\documentclass[sigplan,nonacm]{acmart}
\usepackage{hyperref}
\usepackage{gensymb}
\usepackage[normalem]{ulem}
\usepackage{xcolor}
\usepackage{amsmath,amsfonts}
\usepackage{algorithm}
\usepackage{algorithmic}
\usepackage{graphicx}
\usepackage{textcomp}
\usepackage{enumitem}
\usepackage{listings}
\usepackage{subcaption}
\usepackage{soul}
\usepackage{comment}
\usepackage{tikz}
\usetikzlibrary{positioning}
\usepackage{xspace}
\usepackage{array}
\usepackage{colortbl}
\usepackage{multirow}
\usepackage{makecell}
\usepackage{comment}
\usepackage{graphicx}
\usepackage{subcaption}
\usepackage{booktabs}
\usepackage{tabularx}

\usepackage{booktabs}
\hypersetup{
    unicode=true,
    bookmarksnumbered=true,
    colorlinks=true,
    linkcolor={red!70!black},
    citecolor={red!70!black},
    urlcolor={blue!70!black},
    pdfborder={0 0 0}
}
\usepackage[capitalize,nameinlink]{cleveref}
\usepackage{flushend}
\usepackage{pifont}

\lstdefinestyle{py-acm}{
  language=Python,
  basicstyle=\ttfamily\scriptsize,
  keywordstyle=\bfseries\color{blue!60!black},
  commentstyle=\itshape\color{green!40!black},
  stringstyle=\color{red!55!black},
  numberstyle=\tiny\color{gray},
  backgroundcolor=\color{gray!5},
  showstringspaces=false,
  columns=fullflexible,
  keepspaces=true,
  numbers=left,
  stepnumber=1,
  numbersep=6pt,
  tabsize=2,
  frame=single,
  framerule=0.3pt,
  breaklines=true,
  breakatwhitespace=true,
  captionpos=b,
  aboveskip=0.5em,
  belowskip=0.3em
}

\definecolor{gpuHigh}{RGB}{255,180,80}   
\definecolor{gpuMed}{RGB}{255,210,130}   
\definecolor{gpuLow}{RGB}{255,240,200}   

\definecolor{hgpuHigh}{RGB}{80,120,255}  
\definecolor{hgpuMed}{RGB}{130,170,255}  
\definecolor{hgpuLow}{RGB}{200,220,255}  

\definecolor{grayHigh}{RGB}{160,160,160}   
\definecolor{grayMed}{RGB}{200,200,200}    
\definecolor{grayLow}{RGB}{235,235,235}    

\definecolor{energyHigh}{RGB}{0,120,0}
\definecolor{energyMed}{RGB}{120,200,120}
\definecolor{energyLow}{RGB}{200,240,200}

\definecolor{costHigh}{RGB}{180,0,0}
\definecolor{costMed}{RGB}{240,120,120}
\definecolor{costLow}{RGB}{255,200,200}

\newcommand{\sysname}{TenonAgent\xspace}

\newcommand{\myparagraph}[1]{\vspace{\smallskipamount}\noindent\textbf{#1.\xspace}}

\newcommand*{\rom}[1]{\uppercase\expandafter{\romannumeral #1\relax}}

\newcounter{insightcounter}

\newif\ifshowcomment
\showcommenttrue

\ifshowcomment
    \newcommand{\gohar}[1]{{\color{blue}[GI: #1]}}
    \newcommand{\esha}[1]{{\color{purple}[EC: #1]}}
    \newcommand{\inigo}[1]{{\color{green}[IG: #1]}}
    \newcommand{\haoran}[1]{{\color{orange}[HQ: #1]}}
    \newcommand{\adam}[1]{{\color{lilac}[AB: #1]}}
    \newcommand{\rodrigo}[1]{{\color{magenta}[RF: #1]}}
\else
    \newcommand{\gohar}[1]{\ignorespaces}
    \newcommand{\esha}[1]{\ignorespaces}
    \newcommand{\inigo}[1]{\ignorespaces}
    \newcommand{\haoran}[1]{\ignorespaces}
    \newcommand{\adam}[1]{\ignorespaces}
    \newcommand{\rodrigo}[1]{\ignorespaces}
\fi

\usepackage[most]{tcolorbox}
\tcbset{textmarker/.style={%
        enhanced,
        parbox=false,boxrule=0mm,boxsep=0mm,arc=1.5mm,
        outer arc=1.5mm,left=2mm,right=2mm,top=4pt,bottom=3pt,
        toptitle=1mm,bottomtitle=1mm,oversize}}

\newtcolorbox{simplenoteBox}{colback=white, colframe=black, boxrule=0.2mm, arc=0mm, auto outer arc, boxsep=0mm, left=2mm, right=2mm, top=1mm, bottom=1mm} 
\newtcolorbox{noteBox}{textmarker,
    colback=gray!8!white}

\crefname{insight}{Insight}{Insight}
\newcounter{insight}

\title{\sysname{}: Virtualizing Execution Residency for Memory-Oversubscribed LLM Agents}
\author{Yifan Zhang}
\affiliation{%
  \institution{Zhejiang University}
  \city{Hangzhou}
  \country{China}
}
\email{12451018@zju.edu.cn}

\author{Xinkui Zhao}
\affiliation{%
  \institution{Zhejiang University}
  \city{Hangzhou}
  \country{China}
}
\email{zhaoxinkui@zju.edu.cn}

\author{Ruixin Huang}
\affiliation{%
  \institution{Zhejiang University}
  \city{Hangzhou}
  \country{China}
}
\email{ruixinhuang1@gmail.com}

\author{Lezhuo Xuan}
\affiliation{%
  \institution{Zhejiang University}
  \city{Hangzhou}
  \country{China}
}
\email{693424644@qq.com}

\author{Dianyu Wang}
\affiliation{%
  \institution{Zhejiang University}
  \city{Hangzhou}
  \country{China}
}
\email{dianyuwang61@gmail.com}

\author{Siying Sun}
\affiliation{%
  \institution{Zhejiang University}
  \city{Hangzhou}
  \country{China}
}
\email{12351020@zju.edu.cn}

\author{Xiaoyi Li}
\affiliation{%
  \institution{Zhejiang University}
  \city{Hangzhou}
  \country{China}
}
\email{bettermultiply@gmail.com}

\author{Hengxuan Lou}
\affiliation{%
  \institution{Zhejiang University}
  \city{Hangzhou}
  \country{China}
}
\email{12560049@zju.edu.cn}

\author{Guanjie Cheng}
\affiliation{%
  \institution{Zhejiang University}
  \city{Hangzhou}
  \country{China}
}
\email{chengguanjie@zju.edu.cn}

\author{Naibo Wang}
\affiliation{%
  \institution{Zhejiang University}
  \city{Hangzhou}
  \country{China}
}
\email{wangnaibo@zju.edu.cn}

\author{Chang Liu}
\affiliation{%
  \institution{Zhejiang University}
  \city{Hangzhou}
  \country{China}
}
\email{chang.liu@zju.edu.cn}

\begin{document}

\begin{abstract}
LLM agents execute as long-lived, multi-stage workflows with substantial
resource-demand variation across phases. Provisioning each agent or sandbox
for its lifetime peak leads to low resource utilization, particularly for
memory, whereas aggressive oversubscription risks memory contention,
out-of-memory failures, and costly workflow retries. This tension limits the
concurrency and efficiency of agent serving under constrained host memory.

We present \sysname{}, a runtime system that virtualizes
sandbox residency by decoupling the logical agent's lifetime from its physical execution environment. To manage memory oversubscription, \sysname{} employs a dual-timescale control plane. 
At action boundaries, it combines predictive admission control with a fail-closed, restart-safe contract, dynamically materializing or dematerializing sandboxes based on global memory pressure. 
During active tool execution, a \texttt{sched\_ext}-based kernel gate paces sandbox execution permissions to reduce unexpected concurrent memory growth.
The gate provides feedback control over future execution
opportunities.

Evaluations across diverse agentic workloads show that, under
the tested workloads and parameter settings, \sysname{}
substantially reduces observed OOM failures and achieves a
100\% task success rate in bursts reaching an $8.6\times$
memory oversubscription ratio. Overall, it delivers up to a
$5.5\times$ speedup over unmanaged execution, outperforms
existing agent-level and cgroup-based schedulers by $5\times$
to $37\times$, and provides consistent benefits across both
Docker and Firecracker microVM backends.

\end{abstract}

\maketitle

\pagestyle{plain}

\section{Introduction}
LLM agents are turning cloud AI from sequences of stateless inference requests into long-lived, stateful computations that interleave model reasoning with tool invocation and local execution
~\cite{karpas2022mrkl,schick2023toolformer,shen2023hugginggpt}.
A single task may alternate among remote model calls, shell commands, code
execution, browser interactions, retrieval, and package installation while
carrying state across many steps
~\cite{yao2023reactsynergizingreasoningacting,qin2023toolllm,
openai_function_calling_guide,yao2022webshop,liu2023agentbench,
yang2024sweagent,xu2024crab,xie2024osworld,zhou2024webarena}.

\begin{figure}[t]
    \centering

    \begin{subfigure}[t]{\linewidth}
        \centering
        \includegraphics[width=0.95\linewidth]{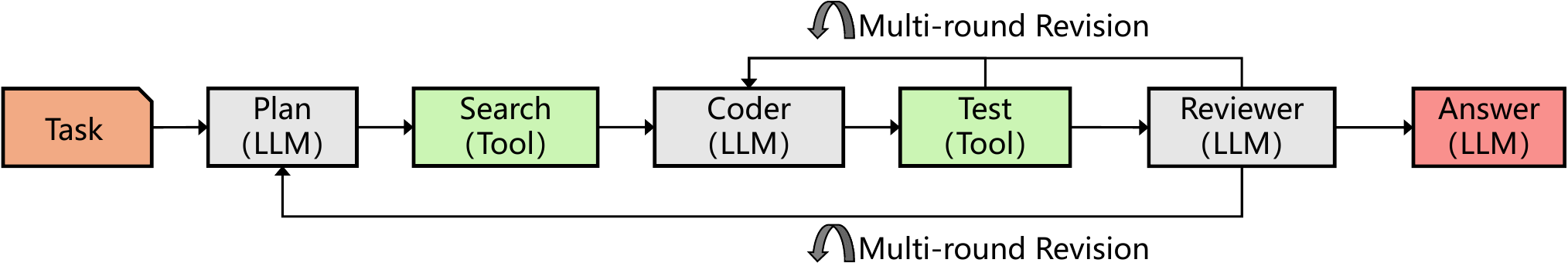}
        \caption{Software-engineering code-generation workflow.}
        \label{fig:swe-workflow}
    \end{subfigure}

    \vspace{0.5em}

    \begin{subfigure}[t]{\linewidth}
        \centering
        \includegraphics[width=0.95\linewidth]{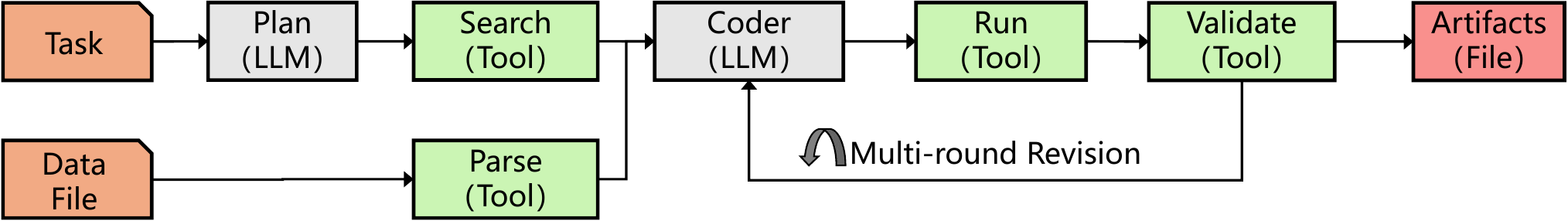}
        \caption{Scientific data-analysis workflow.}
        \label{fig:science-workflow}
    \end{subfigure}

    \caption{Representative agent workflows. Both software-engineering and
    scientific data-analysis agents execute as multi-step, stateful workflows.}
    \label{fig:agent-workflows}
\end{figure}

Agent-serving systems commonly separate orchestration from tool execution.
An external controller maintains the conversation, execution trajectory, and
action metadata~\cite{langchain,langgraph}, while an isolated container or
microVM~\cite{jonas2019cloud,akkus2018sand,agache2020firecracker,
ao2022faasnap} executes local commands, tools, and generated code. The sandbox
provides an isolation and resource-accounting boundary and is commonly retained
across actions, allowing later actions to reuse its files, processes, runtime
state, and cached data. Figure~\ref{fig:agent-workflows} illustrates this
execution model for representative software-engineering and scientific
data-analysis agents.

When many agents share a host, their long-lived sandboxes compete for finite
memory capacity. Provisioning each sandbox for its lifetime peak strands
capacity during less demanding phases, whereas smaller static allocations risk
allocation stalls, swapping, or out-of-memory (OOM) failures when several
sandboxes grow concurrently
~\cite{linuxpsi,gu2017infiniswap,amaro2020fastswap,ruan2020aifm,agentenv2024,
weiner2022tmo}. Moreover, an agent may wait for a remote model or external
service while its sandbox retains runtime state, cached data, and intermediate
artifacts accumulated by preceding actions. Thus, low execution activity does
not imply either a small resident footprint or readily reclaimable memory.

Our profiling
(Section~\ref{sec:profiling-agent-resource-behavior}) reveals that agent memory
demand and execution activity vary across semantically distinct actions.
Waiting-dominated intervals exhibit little local CPU activity and limited
memory growth, whereas tool invocation and code execution can produce bursts of
CPU activity, I/O, and memory allocation. At the same time, a waiting sandbox
may retain substantial memory allocated by earlier actions.
\begin{figure*}[t]
    \centering

    \begin{subfigure}[t]{0.3\textwidth}
        \centering
        \includegraphics[width=\linewidth]{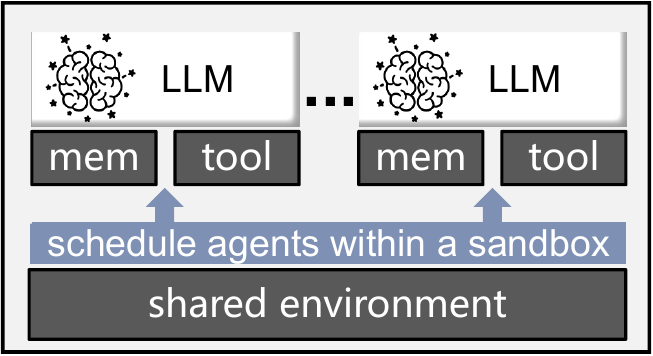}
        \caption{Whole-agent control.}
        \label{fig:scheduling-share}
    \end{subfigure}
    \hfill
    \begin{subfigure}[t]{0.3\textwidth}
        \centering
        \includegraphics[width=\linewidth]{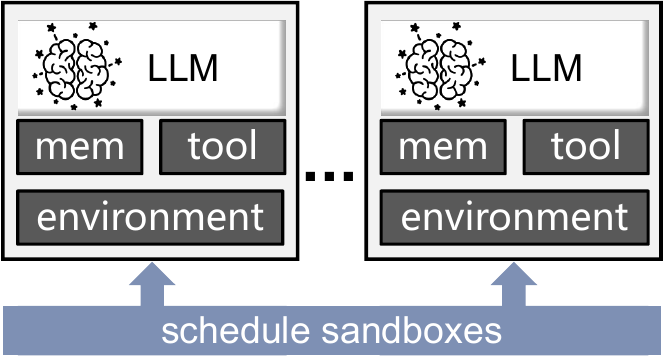}
        \caption{Sandbox-lifetime control.}
        \label{fig:scheduling-split}
    \end{subfigure}
    \hfill
    \begin{subfigure}[t]{0.38\textwidth}
        \centering
        \includegraphics[width=\linewidth]{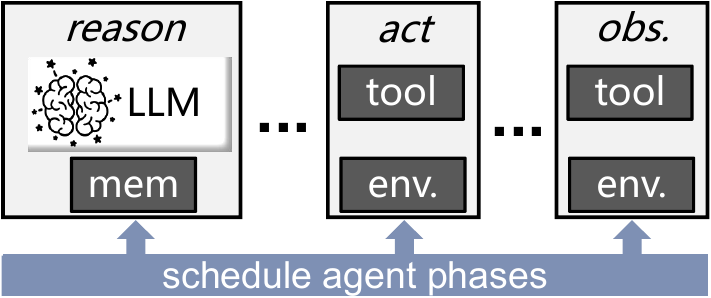}
        \caption{Phase-level control.}
        \label{fig:scheduling-phase}
    \end{subfigure}

    \caption{Resource-control granularity in agent runtimes. Whole-agent
    control makes resource decisions for an entire agent execution, while
    sandbox-lifetime control introduces isolation and accounting but does not
    reconsider resource decisions across actions. Phase-level control
    preserves the sandbox isolation boundary while using action-completion and
    action-ready events as finer-grained control points.}
    \label{fig:scheduling-granularity}
\end{figure*}

This behavior creates an opportunity for semantic resource control.
Before an action executes, the runtime can infer upcoming demand from its type and causal trajectory; upon action completion, it can reconsider whether the sandbox must remain resident.
Action boundaries thus expose natural opportunities to estimate incremental demand before admission and to regulate retained state afterward.
Figure~\ref{fig:scheduling-granularity} contrasts this with existing control granularities.
Whole-agent control treats the entire workflow as a single coarse scheduling unit~\cite{langchain,langgraph}, while sandbox-lifetime control enforces isolation at the container or microVM level~\cite{akkus2018sand,agache2020firecracker,ao2022faasnap} without exposing intra-workflow transitions.
In contrast, action-boundary control decouples logical agent orchestration from physical sandbox residency, enabling fine-grained memory adjustments while preserving sandbox isolation.

Managing memory across these boundaries poses unique challenges distinct from CPU.
Unlike CPU, which can be time-shared or throttled, memory pressure stems from both inactive retained footprints and dynamic allocations during execution, and throttling cannot reclaim allocated space.
Furthermore, while action completion offers a natural reclamation point, dematerializing a sandbox is not always safe: retaining it avoids reconstruction overhead but consumes memory headroom, whereas dematerializing releases memory only when subsequent execution state is durably preserved.
An agent runtime must therefore coordinate action admission, execution growth, and residency, explicitly separating whether a sandbox \emph{can} be safely dematerialized from whether current host pressure makes doing so \emph{beneficial}.

To address this challenge, we present \sysname{}, which virtualizes sandbox
residency by decoupling a logical agent's lifetime from that of its execution
sandbox. While the sandbox transitions between resident and absent states, an
external controller preserves the agent trajectory and action metadata, and a
persistent workspace retains durable execution state. The sandbox can thus be
rematerialized without changing the agent's logical identity.

\sysname{} combines action-boundary decisions with runtime feedback. At action
completion, a fail-closed restart-safe contract determines whether the sandbox
may be dematerialized, after which memory pressure determines whether it should
be retained. At action readiness, \sysname{} predicts incremental demand from
the action and its trajectory, rematerializes the sandbox if necessary, and
admits execution according to available headroom.

During execution, a user-space controller monitors memory pressure and sandbox
growth to correct prediction errors. It updates cgroup-level permissions
enforced by a \texttt{sched\_ext} gate, which delays selected sandbox tasks to
limit further allocation opportunities. Although this mechanism cannot reclaim
existing allocations or immediately block new ones, it forms a closed loop with
action-boundary control: user space manages semantic policy, while the kernel
enforces sandbox-wide execution decisions.

We implement \sysname{} for Docker containers and Firecracker microVMs and
evaluate it on three LLM-agent workloads under constrained memory. We compare
against eager Linux, static low- and high-concurrency configurations, an
agent-level reservation scheduler~\cite{kang2026thunderagent}, and AgentCgroup~\cite{zheng2026agentcgroup}. We measure makespan, P50
workflow latency and peak memory, and OOM and timeout
events; failed workflows are re-enqueued so that retry costs are included.

This paper makes the following contributions:

\begin{itemize}
    \item We introduce \emph{sandbox residency virtualization}, which decouples
    an LLM agent's logical lifetime from its physical execution sandbox. With
    restart-safe action boundaries, eligible idle sandboxes can be
    dematerialized and deterministically reconstructed without breaking logical
    identity or task progression.

    \item We design a two-timescale host-memory control policy that combines
    pressure-aware residency management, predictive action admission, and
    runtime execution pacing. This reduces the impact of demand-prediction
    errors and varying host-memory contention without static per-sandbox
    reservations.

    \item We implement \sysname{} for Docker and Firecracker and evaluate it on
    SWE-bench, ScienceAgentBench, and Orchard RL Rollout. Under the tested
    workloads and parameter settings, \sysname{} substantially reduces observed
    OOM failures, achieves a 100\% task success rate in bursts reaching an
    \textbf{8.6$\times$ memory oversubscription ratio}, and delivers up to a
    \textbf{5.5$\times$ speedup} over unmanaged execution.
\end{itemize}

\section{Background and Motivation}

\subsection{Agent Workflows Management}
Representative agent workflows (e.g., Figure~\ref{fig:agent-workflows}) interleave model reasoning with environment actions. Outputs from each step can dynamically alter the environment through files, dependencies, or runtime processes. Consequently, an agent workflow operates as a long-lived, stateful program rather than a sequence of stateless model requests. Resource demand fluctuates substantially across these steps. For instance, waiting on external LLM calls incurs negligible local activity, whereas local tool executions can trigger sharp bursts in CPU, I/O, and memory allocations. 

In production deployments, agents typically execute within isolated sandboxes (e.g., containers~\cite{miell2019docker, zheng2026agentcgroup} or microVMs~\cite{agache2020firecracker,cheng2026llminsandbox,dong2026deltabox}). Existing orchestration platforms, such as Kubernetes~\cite{kubernetes}, manage resources by applying static limits at the container or pod level, enforced by Linux cgroups~\cite{rosen2013resource,rosen2016namespaces}. Under this model, the execution sandbox and its resource allocations persist for the entire lifetime of the workflow, independent of the dynamic phase changes occurring inside the application.

\subsection{Empirical Motivation}
\label{sec:profiling-agent-resource-behavior}

\begin{figure}[t]
    \centering
    \includegraphics[width=\linewidth]
        {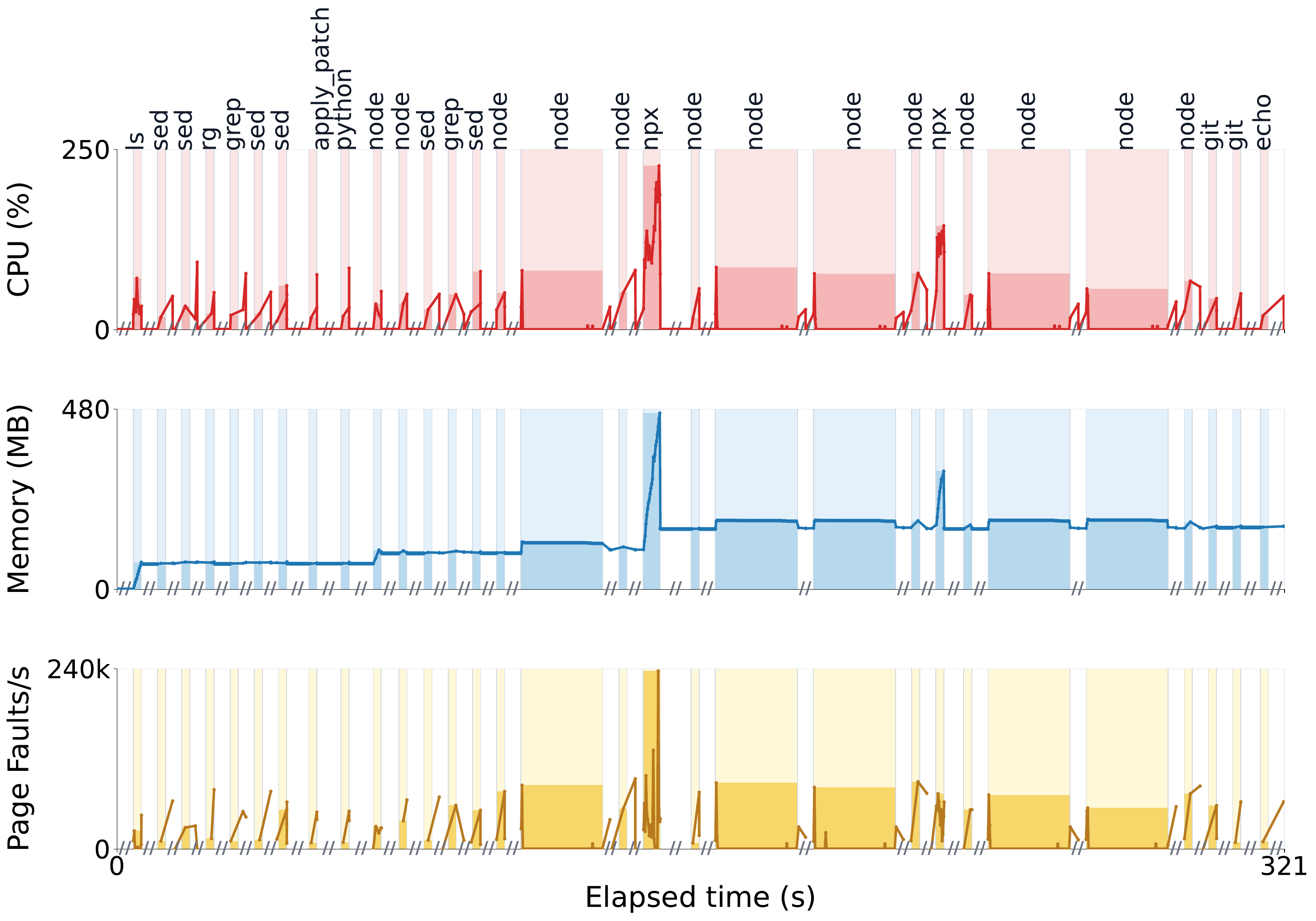}
    \caption{Resource dynamics of a representative SWE-agent execution,
    aligned with semantic events in its execution trajectory.}
    \label{fig:profiling}
\end{figure}

\textbf{Experiment 1: Profiling Resource Dynamics.}
We profile representative SWE-agent~\cite{yang2024sweagent} executions and
collect user-space and OS-level signals, including CPU utilization, resident
memory, page-fault activity, I/O activity, and memory pressure stall
information (PSI). We align these measurements with semantic events in the
execution trajectory, including model invocations, tool calls, local commands,
and result validation. Figure~\ref{fig:profiling} shows a representative
execution trace together with its action annotations.

\textbf{Observation 1: Semantic Phases Delineate Distinct Resource Regimes.}
The profiled traces reveal that resource demand fluctuates drastically, and these shifts align strictly with semantic transitions rather than physical time intervals. During the execution of an action, local computation produces sharp bursts in CPU activity, page faults, and memory allocation. However, once the action completes and the agent enters a waiting phase (e.g., awaiting an LLM response), new allocations cease and CPU utilization drops to near zero. Despite this inactivity, the memory accumulated by preceding actions typically remains resident—for instance, plateauing at roughly 150\,MB after a 469\,MB burst in our representative trace. These distinct regimes suggest that semantic phase boundaries (e.g., action-completion and action-ready events) provide natural, workload-aware intervention points to reconsider resource allocation, offering a more precise control granularity than fixed time windows.

\begin{figure}[t]
    \centering
    \includegraphics[width=\linewidth]
        {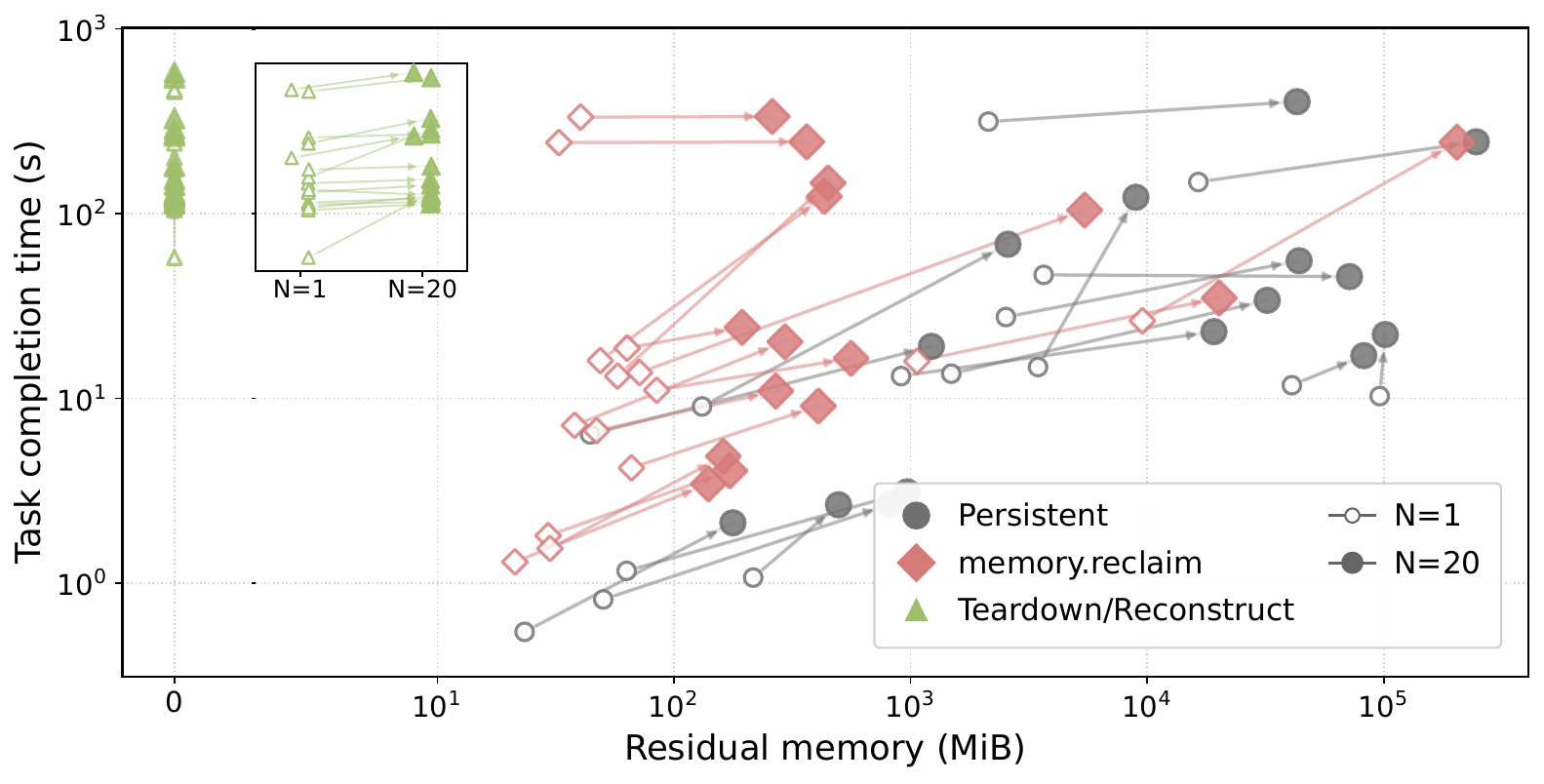}
    \caption{Aggregate residual memory versus task completion time under
    different boundary policies. Open and filled markers denote $N=1$ and
    $N=20$, respectively. Each line connects the two concurrency levels for
    the same benchmark and policy, rather than representing a continuous
    trade-off curve.}
    \label{fig:experiment2}
\end{figure}

\textbf{Experiment 2: Boundary-Policy Trade-offs under Concurrency.}
We compare memory-management policies applied at the same boundaries between action-execution phases. Retaining a sandbox preserves warm state and avoids reconstruction overhead, but continues to occupy memory. Teardown releases nearly all sandbox memory, but subsequent execution incurs rematerialization and re-warming costs.
For each policy, we simultaneously launch either one workflow ($N=1$) or twenty workflows ($N=20$). 
All policies use the same replayed trajectories and reconstruct semantically equivalent environments. We measure task completion time and aggregate residual memory after boundary processing. 
Aggregate memory captures the host-level footprint accumulated across concurrent workflows. Figure~\ref{fig:experiment2} shows the resulting memory--latency trade-offs.

\textbf{Observation 2: Concurrency Shifts the Memory--Latency Trade-off.}
The results demonstrate that no single boundary policy is optimal across all operating conditions. At low concurrency ($N=1$), retention-oriented policies significantly reduce task completion time by avoiding the substantial overhead of sandbox reconstruction and environment re-warming, albeit at the cost of a larger residual memory footprint. However, as concurrency scales to $N=20$, the retention policy causes retained memory to accumulate rapidly across all idle sandboxes. This aggregate footprint creates severe host-level memory contention, which inflates the completion time and entirely erases the low-concurrency latency advantage. Conversely, the teardown policy releases nearly all memory at action boundaries, maintaining a consistently small residual footprint. While teardown incurs a baseline rematerialization penalty, its completion time remains much less sensitive to increased concurrency. Consequently, boundary management must be adaptive: favoring retention when memory headroom is abundant, and shifting to proactive reclamation as aggregate memory pressure rises.

\textbf{Implications.} Together, these observations motivate semantic and load-adaptive memory management. Action boundaries provide practical decision points, but reclamation must account for restart safety and current system conditions.
Moreover, boundary policies control memory retained between actions, not memory growth after execution begins. Effective oversubscription therefore requires jointly managing sandbox residency, action admission, and execution-time memory growth.

\section{\sysname{} Design}
\label{sec:design}

\subsection{Overview}
\label{sec:design:overview}

\sysname{} is built around a separation between logical agent identity and
physical sandbox residency. A logical agent is defined by the external
controller state and durable workspace that persist across actions, whereas
the sandbox is the physical execution substrate that holds ephemeral state such
as processes, runtime caches, services, and live connections. By separating
these two notions, \sysname{} can keep an idle sandbox warm when memory is
available, or safely dematerialize it under memory pressure without changing
the identity of the agent.

\begin{figure}[t]
    \centering
    \includegraphics[width=\linewidth]{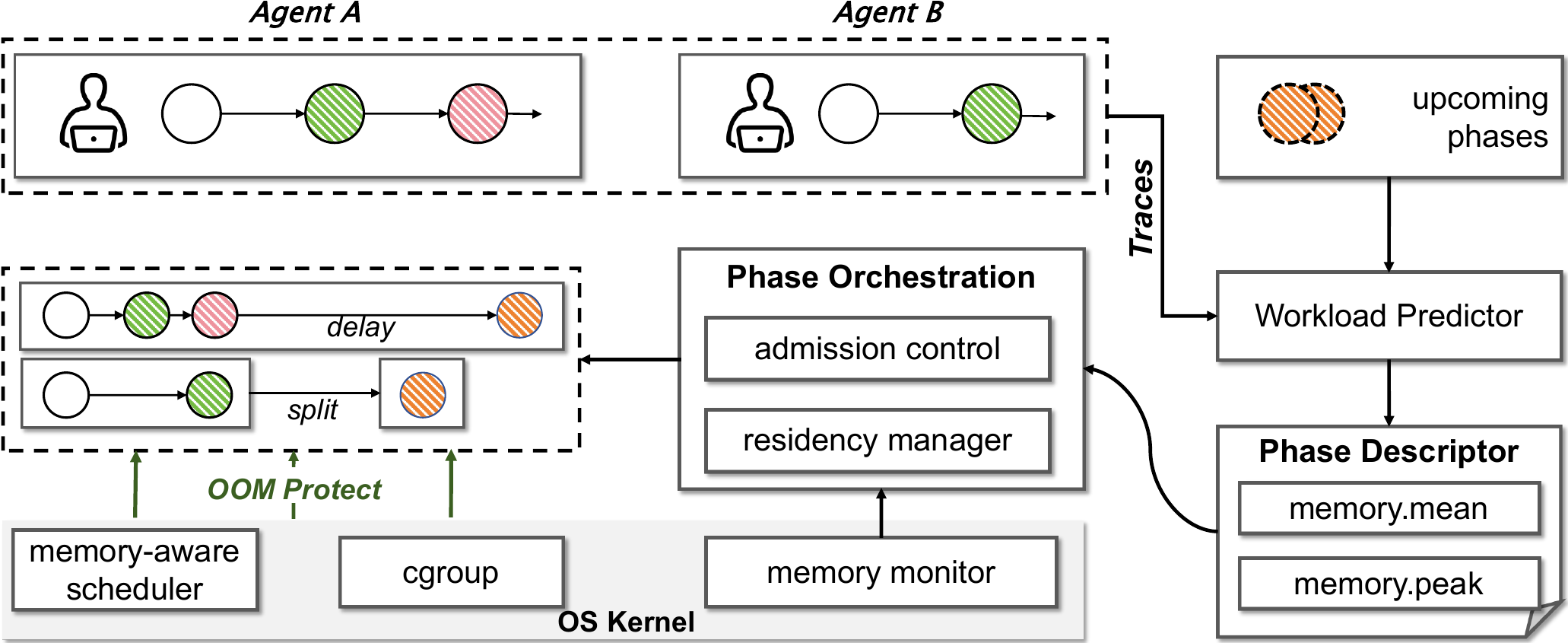}
    \caption{\sysname{} architecture for agent sandbox residency virtualization.}
    \label{fig:architecture}
\end{figure}

Figure~\ref{fig:architecture} illustrates this architecture, which is governed by a \textbf{dual-timescale control plane}. 
At coarse-grained action boundaries, a user-space orchestrator leverages demand predictions to admit new execution phases, dynamically managing idle sandbox residency based on global memory pressure. 
At the fine-grained execution timescale, a \texttt{sched\_ext}-based kernel gate monitors instantaneous memory growth, pacing CPU permissions to throttle sandboxes that deviate from their admitted limits. 
Ultimately, \sysname{} couples proactive demand prediction to unlock high concurrency with reactive kernel enforcement to help maintain system safety under extreme oversubscription.

\subsection{Global Memory-Pressure State}
\label{sec:design:pressure-state}

Unlike CPU time, memory remains occupied even when execution pauses. Memory
control must therefore manage both memory retained by resident sandboxes and
additional memory that active phases may allocate. Because reclaiming retained
memory takes time, \sysname{} must respond before active execution exhausts the
remaining headroom. It captures this global condition using three pressure
states---\textsc{green}, \textsc{yellow}, and \textsc{red}---denoted by $Q(t)$.

The user-space controller determines $Q(t)$ by monitoring the available memory
of the workload-domain parent cgroup and estimating the remaining growth of
already-admitted phases. \textsc{Green} indicates that sufficient headroom
remains for both current occupancy and expected growth. \textsc{Yellow}
indicates that the system is still safe but its margin is narrowing, requiring
preventive reclamation. \textsc{Red} indicates that little headroom remains or
that active phases may consume the remaining capacity, requiring immediate
restrictions on further memory growth. Accounting for predicted growth allows
\sysname{} to react before current memory usage alone reveals critical
pressure.

\newcommand{\cmark}{\ding{51}}
\newcommand{\xmark}{\ding{55}}

\begin{table}[t]
\centering
\small
\renewcommand{\arraystretch}{0.9}
\setlength{\tabcolsep}{6pt}
\caption{System policies in each pressure state
($\sim$: selective throttling).}
\label{tab:pressure-matrix}
\begin{tabular}{lccc}
\toprule
\textbf{Mechanism / Policy}
    & \textsc{Green}
    & \textsc{Yellow}
    & \textsc{Red} \\
\midrule
Retain idle sandboxes
    & \cmark & \xmark & \xmark \\
Predictive rematerialization
    & \cmark & \xmark & \xmark \\
Admit new execution
    & \cmark & \cmark & \xmark \\
Unrestricted memory growth
    & \cmark & $\sim$ & \xmark \\
\bottomrule
\end{tabular}
\end{table}

As summarized in Table~\ref{tab:pressure-matrix}, $Q(t)$ coordinates two complementary control paths. User-space sandbox lifecycle management (\S\ref{sec:design:user-scheduler}) reclaims retained memory from idle sandboxes, while kernel-enforced execution control (\S\ref{sec:design:kernel-enforcement}) limits further memory growth in active ones. The user-space controller determines the overall policy, and the kernel promptly enforces the resulting execution permissions.

\subsection{Action Descriptors and Resource Prediction}
\label{sec:design:phase-descriptor}

\paragraph{Action descriptors.}
For each upcoming action $\phi_i$, \sysname{} constructs a descriptor
\[
D_i =
\left\langle
c_i,\,
\widehat{M}^{avg}_i,\,
\widehat{M}^{peak}_i,\,
\widehat{H}_i,\,
p_i
\right\rangle ,
\]
where $c_i$ denotes the action class,
$\widehat{M}^{avg}_i$ and $\widehat{M}^{peak}_i$ are its predicted average and
peak cgroup footprints, $\widehat{H}_i$ captures heavy-memory risk, and $p_i$
represents scheduling priority or latency sensitivity. Together, these fields
model the action's expected resource demand and scheduling requirements, serving 
as inputs to admission and execution control.

\begin{figure}[t]
    \centering
    \includegraphics[width=\linewidth]{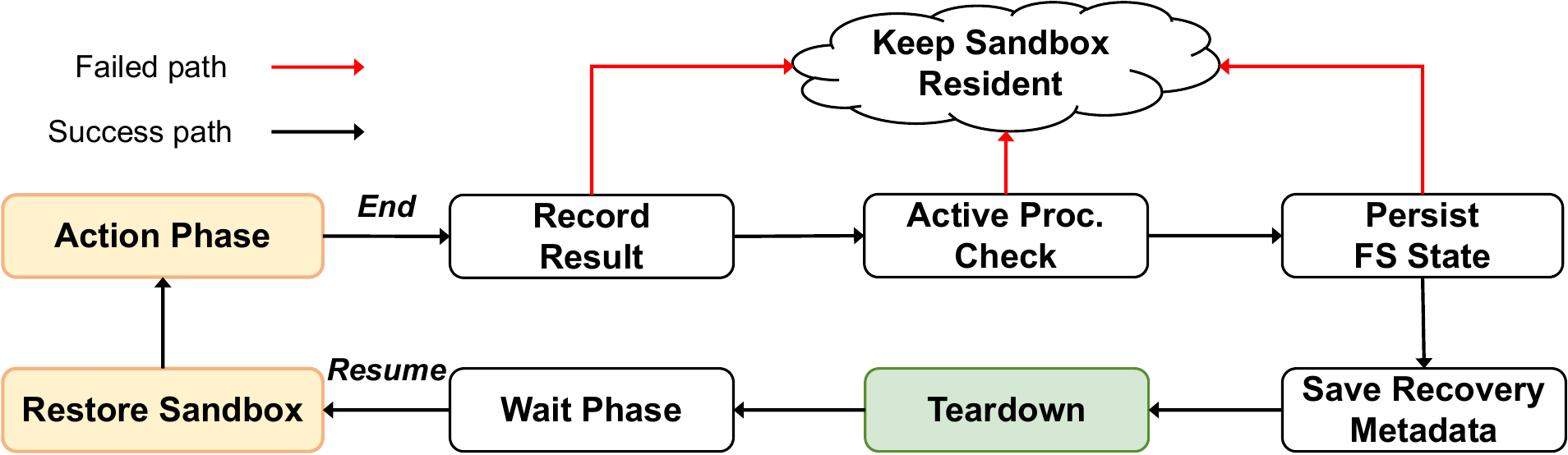}
    \caption{Restart-safe teardown and restoration.
    Preparation failures retain the resident sandbox.}
    \label{fig:restart-safe-teardown}
\end{figure}
\paragraph{Restart-safe recoverability.}
Separately, \sysname{} labels the boundary preceding $\phi_i$ as either
\texttt{restartable} or \texttt{resident\_only}. \sysname{} supports
restart-safe recovery under a restricted action execution model rather than
inferring restartability for arbitrary programs. \sysname{} does not instrument
or rewrite the agent to make an arbitrary execution restartable; instead, it
checks whether the boundary exposed by the existing agent execution satisfies
this restricted restart-safe contract. A boundary is \texttt{restartable} only
when the completed action has externalized its logical state, left no required
state exclusively in live processes or connections (e.g., persistent shells or
background services), and persisted durable artifacts for deterministic
reconstruction. Figure~\ref{fig:restart-safe-teardown} shows the protocol when sandbox release is
selected. The controller blocks new action dispatch throughout teardown
preparation and determines eligibility through deterministic runtime checks,
without LLM-based judgment. Teardown is permitted only after a validated
recovery record has been committed, linking the logical agent state, filesystem
artifact versions, and runtime configuration to the completed action boundary.
This protocol is strictly fail-closed: any unsupported execution or preparation
failure defaults the boundary to \texttt{resident\_only}. As a correctness
constraint, memory-pressure policies may still throttle a
\texttt{resident\_only} sandbox, but cannot tear it down. Upon
rematerialization, restoration reconstructs the recorded environment and state;
it neither replays the completed action nor restores arbitrary process memory.
We detail these execution constraints and recovery procedures in Appendix.

\subsection{User-Space Boundary-Scale Scheduling}
\label{sec:design:user-scheduler}

\begin{figure}[t]
    \centering
    \includegraphics[width=\linewidth]{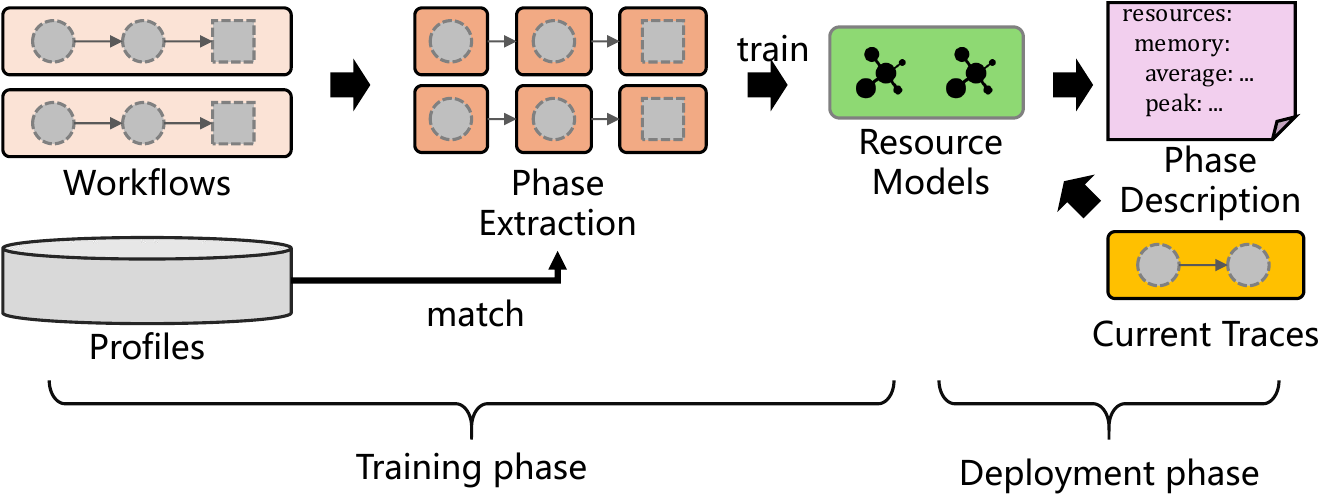}
    \caption{Pipeline for constructing action-level resource predictors and
    generating causal predictions at runtime.}
    \label{fig:resource-predictor}
\end{figure}

\paragraph{Causal resource prediction.}
Figure~\ref{fig:resource-predictor} shows the training and runtime prediction
pipeline. Before action $\phi_i$ executes, \sysname{} constructs a causal
feature vector from its action context, summaries of completed actions in the
same task, and TF--IDF features derived from the prompt, current action, and
recent trajectory. The predictor uses only information available before
$\phi_i$ begins; it cannot access the action's realized resource usage,
outcome, failure state, or any future action.

Separate models predict average memory, peak memory, and heavy-memory risk,
allowing the scheduler to use different levels of conservatism for normal
accounting, peak-aware admission, and high-risk actions. After $\phi_i$
completes, its measured average footprint, peak footprint, and heavy-memory
label are appended to the task-local history and become available only to
later predictions. Training samples are constructed by aligning resource
profiles with extracted action phases, and task-level train/validation splits
prevent leakage across correlated actions.

At semantic action boundaries, \sysname{} coordinates action admission and
idle-sandbox residency. Admission uses action descriptors to maintain a logical
memory water level based on the predicted charges of admitted actions, whereas
residency control uses the global pressure state $Q(t)$ to retain or reclaim
sandboxes during wait phases. This separation enables admission to anticipate
near-term demand while residency control responds to system-wide pressure.

\paragraph{Water-level-aware action admission.}

\begin{figure}[t]
    \centering
    \includegraphics[width=0.95\linewidth]{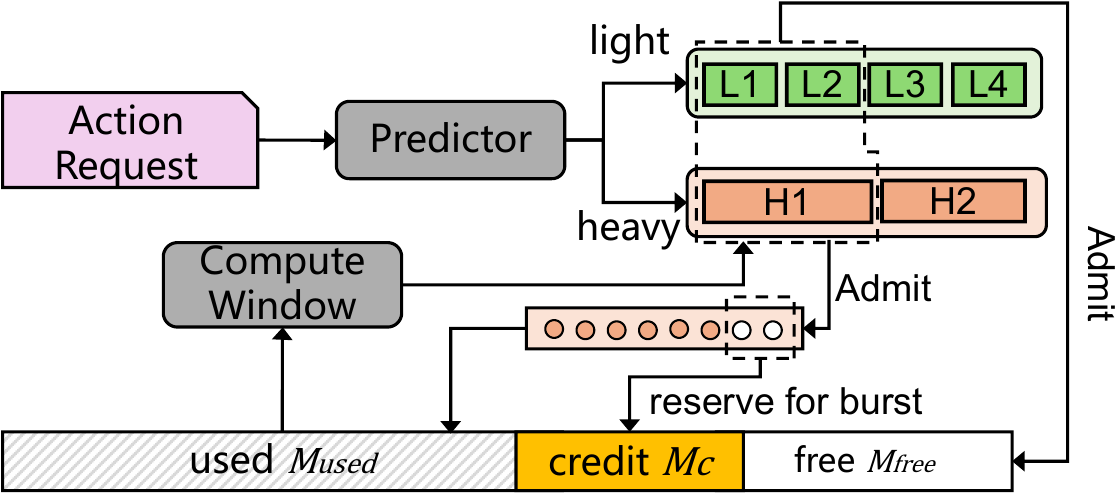}
    \caption{Water-level-aware admission with light and heavy queues sharing
    one logical memory budget.}
    \label{fig:dual-queue}
\end{figure}

\sysname{} handles actions with heterogeneous memory demands. Strict FIFO
can cause head-of-line blocking when the oldest action does not fit, while
always prioritizing small actions can starve memory-intensive ones.
\sysname{} therefore combines water-level-aware memory accounting,
logical and physical admission checks, and the light/heavy queues shown
in Fig.~\ref{fig:dual-queue}. The queues share a logical memory budget
and differ only in scheduling priority.

For each action $i$, \sysname{} assigns a logical memory charge
$\widetilde M_i$ by interpolating between its predicted average
($\widehat M_i^{\mathrm{avg}}$) and peak
($\widehat M_i^{\mathrm{peak}}$) demands. The interpolation factor
depends on $M_{\mathrm{admitted}}$, the total logical charge of
currently admitted actions:
\[
\widetilde M_i =
(1-\alpha)\widehat M_i^{\mathrm{avg}}
+\alpha\widehat M_i^{\mathrm{peak}},
\qquad
\alpha =
\left[\frac{M_{\mathrm{admitted}}}{M_{\mathrm{window}}}\right]_{0}^{1}.
\]
Here, $M_{\mathrm{window}}$ is a configured capacity threshold
(e.g., 6\,GB), and $[\,\cdot\,]_{0}^{1}$ clamps its argument to $[0,1]$.
Thus, the charge moves from the predicted average toward the predicted
peak as the admitted logical footprint approaches $M_{\mathrm{window}}$.
The charge is fixed at admission and released when the action completes,
regardless of whether its sandbox remains resident. Among admissible actions, \sysname{} normally prioritizes the light queue
to avoid leaving capacity unused while a heavy action waits. To prevent
starvation, the oldest heavy action accrues protected logical credit at
rate $\beta$; light actions cannot consume this credit. As running actions
complete, the credit protects an increasing share of the logical budget
for that heavy action. It affects scheduling priority only.
\paragraph{Wait-phase residency and rematerialization.}
Residency control begins when an action completes. A sandbox may be reclaimed
during the wait phase only if the completed boundary is \texttt{restartable};
otherwise, it remains resident. Under \textsc{green}, \sysname{} retains idle
sandboxes by default and reclaims one only when the predicted wait amortizes
teardown and rematerialization. Under \textsc{yellow} or \textsc{red}, it
prioritizes memory recovery and reclaims restartable sandboxes in decreasing
order of measured working set.

Rematerialization is triggered when the next action becomes ready. Under
\textsc{green}, \sysname{} may also perform lower-priority predictive
rematerialization. Under \textsc{yellow}, speculative work is disabled, but
demand-driven rematerialization may proceed if it passes the physical gate and
obtains a lifecycle slot. Under \textsc{red}, no new rematerialization starts;
queued requests are reconsidered after pressure falls. Rematerialization
restarts the execution service, restores the persistent workspace and
controller metadata, checks readiness, and refreshes the sandbox's cgroup
identity.

Concretely, at $b_i^{done}$, \sysname{} records the result, determines
\textsc{RestartSafe}$(i)$, releases action $i$'s admission charge, and applies
the residency policy without requiring the next action's descriptor. At
$b_{i+1}^{ready}$, it rematerializes a non-resident sandbox according to the
current pressure state. Reconstruction is included in physical accounting only
while active, and the waiting action holds no admission charge. Once the
sandbox is ready, the action competes for logical admission and passes the
physical gate before execution; runtime permission is subsequently reevaluated
at each control update.

\begin{algorithm}[t]
\caption{Runtime execution permission control}
\label{alg:kernel-enforcement}
\small
\begin{algorithmic}[1]
\REQUIRE Live sandboxes $\mathcal{C}$, time $t$, available memory $M$,
         demands $d_c$, and retained controller state
\ENSURE Permitted set $\mathcal{A}$ and updated controller state

\STATE $H\leftarrow\max(M-S,0)$;
       $R_{\mathrm{prev}}\leftarrow
       \sum_{c\in\mathcal{A}_{\mathrm{prev}}\cap\mathcal{C}}d_c$
\STATE Determine pressure $p$ from $M$ and $R_{\mathrm{prev}}$,
       giving \textsc{red} priority
\STATE Update $b$ and the Green stability timer
\IF{$p=\textsc{green}$}
    \STATE $\mathcal{A}\leftarrow\mathcal{C}$
\ELSE
    \STATE $\mathcal{E}\leftarrow
        \{c\in\mathcal{C}\mid
        \tau_c\text{ unset}\ \lor\ t-\tau_c\ge D_{\min}\}$
    \STATE Order $\mathcal{E}$: overdue first ($D_{\max}$),
           then the prescribed tie-breaking rules
    \STATE $K\leftarrow b$ if $p=\textsc{yellow}$, otherwise $1$;
           $(\mathcal{A},R,n)\leftarrow(\emptyset,0,0)$
    \FORALL{$c\in\mathcal{E}$ in order}
        \STATE $x_c\leftarrow\mathbf{1}[d_c>0]$
        \IF{$R+d_c\le H$ \AND $n+x_c\le K$}
            \STATE $\mathcal{A}\leftarrow\mathcal{A}\cup\{c\}$;
                   $R\leftarrow R+d_c$; $n\leftarrow n+x_c$
        \ENDIF
    \ENDFOR
\ENDIF
\STATE Publish \textsc{allow} for $\mathcal{A}$ and
       \textsc{deny} for $\mathcal{C}\setminus\mathcal{A}$
\STATE Clear timers for allowed sandboxes; start unset denial
       timers at $t$ and retain existing ones
\STATE Change denied Growing sandboxes to PausedGrowing
\STATE $\mathcal{A}_{\mathrm{prev}}\leftarrow\mathcal{A}$;
       $p_{\mathrm{prev}}\leftarrow p$
\end{algorithmic}
\end{algorithm}

\subsection{Runtime Execution Control}
\label{sec:design:kernel-enforcement}

\sysname{} controls memory pressure before it turns into an OOM event. The key
observation is that the scarce resource during execution is often not a
sandbox's current footprint, but its near-future memory growth. A sandbox with a
stable working set can usually keep running safely, while a smaller but
aggressively allocating sandbox may quickly consume the remaining headroom.
Thus, \sysname{} schedules positive memory growth as a runtime resource,
analogous to how congestion control regulates sending rate rather than reacting
only after a queue has filled.

Algorithm~\ref{alg:kernel-enforcement} summarizes the control loop. Each
iteration estimates per-sandbox growth from cgroup-accounted memory samples,
classifies global pressure as \textsc{green}, \textsc{yellow}, or \textsc{red},
and publishes the corresponding allow/deny decisions to the scheduler. In
\textsc{green}, all live sandboxes run. Under pressure, \sysname{} admits only
a bounded number of growing sandboxes whose predicted growth fits the remaining
headroom; in \textsc{red}, this bound collapses to a single headroom-checked
probe.

\paragraph{Growth-aware admission.}
The controller distinguishes sandboxes that are growing from those that appear
resident or cooling. This distinction is used only for execution control: a
sandbox with no predicted positive growth does not consume a growth slot, while
a sandbox with positive predicted growth does, even if its current footprint is
small. This design prevents the scheduler from being biased toward small but
rapidly allocating sandboxes and instead prioritizes executions that are less
likely to increase memory pressure.

Admission in Yellow follows a congestion-control rule. Rather than fixing a
single conservative concurrency limit, \sysname{} uses additive increase and
multiplicative decrease to adapt the number of concurrent growth contributors.
When the system remains stable, the controller cautiously increases the growth
budget; when pressure worsens, it cuts the budget back. This lets the system
probe for available memory parallelism while reacting quickly to congestion,
mirroring the way AIMD network controllers search for safe bandwidth.

\paragraph{Red-mode probing.}
When memory pressure reaches Red, \sysname{} stops treating concurrency as the
goal. The objective becomes controlled progress without allowing a burst of new
allocations. Red therefore permits at most one growing sandbox at a time, and
only if its predicted growth fits in the current headroom. This sandbox acts as
a probe: it may continue, be replaced, or be denied again at the next control
iteration. The one-probe rule also prevents starvation from becoming an
unbounded memory risk. Long-denied sandboxes are prioritized for
reconsideration, but they are not admitted unconditionally; every admission
still has to satisfy the current headroom and growth-slot constraints.

\paragraph{Actuation through sched\_ext.}
The controller enforces these decisions through \texttt{sched\_ext}. Each
sandbox is placed in a dedicated cgroup, and the controller publishes an
allow/deny decision for each cgroup to a BPF map. Denied runnable tasks are
held in a waiting dispatch queue, while allowed tasks are eligible for normal
CPU dispatch. This makes memory control proactive: instead of waiting for the
kernel OOM path or relying only on post-hoc reclamation, \sysname{} can slow
down memory allocation by controlling when allocating sandboxes receive CPU
time. Execution control is deliberately separated from reclamation. The scheduler
limits concurrent predicted growth, whereas restart-safe teardown handles cases
where memory must be actively reclaimed. If Red pressure persists and no
sandbox can be safely torn down, the controller continues to re-evaluate
headroom and to run only headroom-checked probes. The projection is a scheduling
criterion rather than a hard capacity proof: growth estimates may be wrong, and
memory can change before the next control iteration.

\section{Implementation}
\label{sec:implementation}

We implement \sysname{} as a user-space controller with Docker and Firecracker sandbox backends. The resource predictor uses a random forest regression model with semantic, scope, and resource-history features to predict memory usage. Runtime execution control is implemented using \texttt{sched\_ext} and eBPF. 
The Docker and Firecracker backends expose a common lifecycle interface over
backend-specific shutdown, and restore operations.
 The backend reports readiness
to the controller only after the sandbox is available for action execution. 
Sandbox memory usage and pressure are collected through cgroup v2. The
controller stores per-sandbox execution permissions in a cgroup-keyed BPF map,
which the \texttt{sched\_ext} scheduler consults when dispatching sandbox
tasks.

\section{Evaluation}
\label{sec:eval}

\subsection{Experimental Setup}

\paragraph{Evaluation modes.}
We use two complementary evaluation modes: trajectory replay and online
LLM-driven execution. Unless otherwise stated, we use replay for repeatable
comparisons. Replay trajectories are collected from GPT-5.5-driven agent runs;
replay then re-executes the recorded tool calls in real sandboxes, reproduces
recorded waiting times, and follows fixed action sequences rather than adapting
to failures. Each replay experiment is repeated three times with the same
trajectories, submission schedule, and host-memory budget, and we report the
average unless otherwise noted.

\myparagraph{Workloads}
We evaluate our system on representative stateful agent workloads spanning software engineering, scientific research, and reinforcement-learning training:
\begin{itemize}[leftmargin=*]
    \item \textbf{SWE-bench~\cite{yang2024sweagent}:} Represents complex, multi-turn software engineering tasks where agents navigate repositories, compile code, and execute tests.
    \item \textbf{ScienceAgentBench~\cite{chen2024scienceagentbenchrigorousassessmentlanguage}:} Targets data-intensive scientific research workloads that demand significant memory for data processing and tool execution.
    \item \textbf{Orchard RL Rollout~\cite{peng2026orchard}:} Represents reinforcement-learning training environments, generating synchronized bursts of interactive rollout trajectories to stress-test concurrent execution.
\end{itemize}
\begin{figure*}[t]
    \centering
    \includegraphics[width=\linewidth]
        {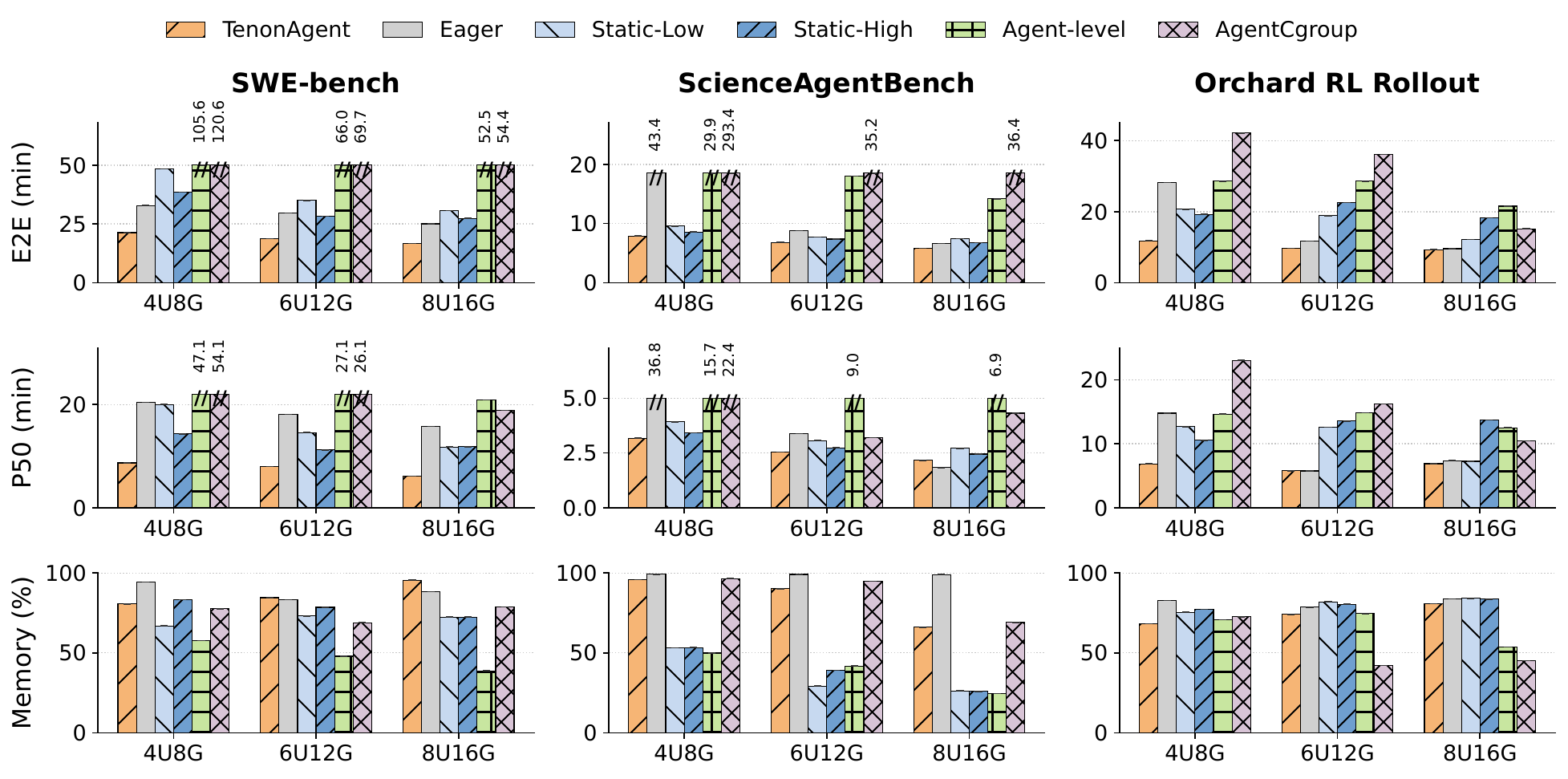}
    \caption{End-to-end latency (top), median task latency (middle), and peak
    memory utilization (bottom) across three workloads and three machine
    configurations. Numbers above truncated bars report their actual values.}
    \label{fig:main-e2e}
\end{figure*}

\myparagraph{Platform}
Our experiments run in virtual machines hosted on an Intel Core i9-14900K
machine with 128\,GiB of DRAM. Each VM is provisioned with the CPU and memory
configuration reported in the corresponding figure (e.g., 4 vCPUs and
8\,GiB, denoted 4U8G). The VMs run Ubuntu 24.04.4 LTS with Linux
7.0.0-31-generic. We use Docker 29.7.2 for container experiments and
Firecracker v1.12.1 for microVM experiments. Agent executions are implemented
using the Mini-SWE-Agent framework~\cite{yang2024sweagent,tan2026automatically}. Pressure and resident-memory measurements
are collected from the workload-domain cgroup, for which swap is disabled.

\myparagraph{Baselines}
We compare \sysname{} with four baselines:

\begin{itemize}[leftmargin=*]
    \item \textbf{Eager:} runs sandboxes with the default Linux memory management and no resource control.

    \item \textbf{Fixed-concurrency:} statically limits concurrent agents, evaluated under conservative (\textsc{Static-Low}) and aggressive (\textsc{Static-High}) bounds.

    \item \textbf{Agent-level peak-reservation scheduler:} implements a lifetime peak-reservation baseline inspired by the reservation-based admission policy of ThunderAgent~\cite{kang2026thunderagent}. In this baseline, each agent reserves its offline-profiled peak host-memory footprint for the lifetime of its container.

    \item \textbf{AgentCgroup~\cite{zheng2026agentcgroup}:} uses
    tool-call-aligned hierarchical cgroups and in-kernel eBPF enforcement to adaptively control agent CPU and memory resources.
\end{itemize}

\myparagraph{Metrics}
Across all configurations, workflows encountering an OOM event or
sandbox timeout are re-enqueued for up to five retries after the initial
attempt. All workflows complete within this limit. We report workload
makespan and P50 workflow latency, both including retries, as well as
peak aggregate sandbox memory, OOM events, and sandbox timeouts. We
also report the attempt-level failure rate, computed as failed attempts
divided by total attempts; its denominator may vary across
configurations because of retries.

\subsection{End-to-End Performance}
\label{sec:eval:e2e}

\begin{figure}[t]
    \centering
    \includegraphics[width=\linewidth]
        {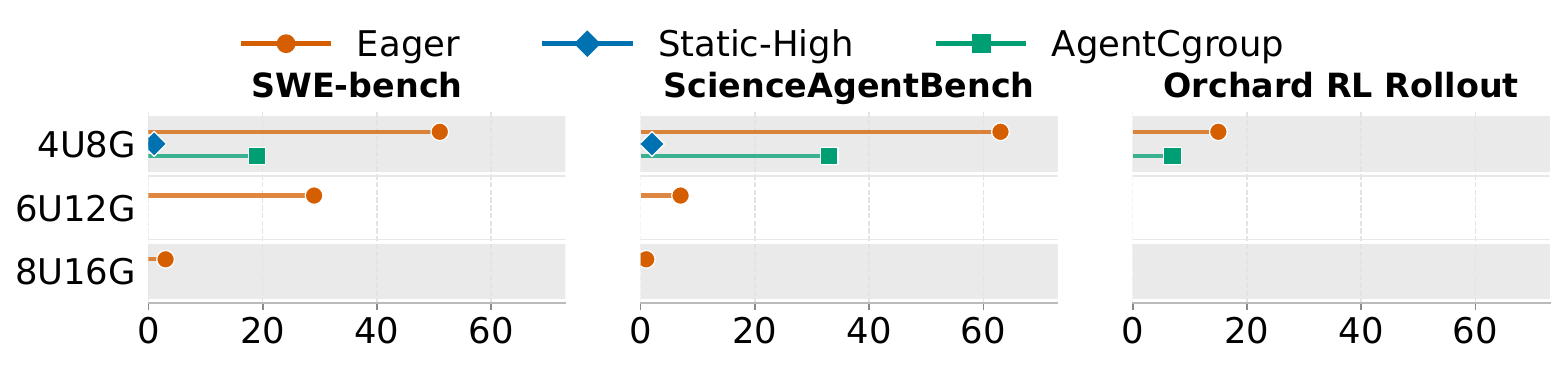}
    \caption{Numbers of tasks that fail because of out-of-memory errors or timeouts. Methods and configurations with no observed failures are omitted.}
    \label{fig:failure-counts}
\end{figure}

Figure~\ref{fig:main-e2e} evaluates all methods under a fixed offered load (7.2 QPS over a 10-second submission window) across varying host capacities. \sysname{} consistently achieves the lowest E2E time throughout this range. Under tightly constrained settings (e.g., 4U8G), \sysname{} dramatically outpaces Eager (Unlimited), delivering up to a \textbf{$5.5\times$ speedup} in batch completion on ScienceAgentBench (7.9 vs. 43.4 minutes) and a \textbf{$2.4\times$ speedup} on Orchard RL Rollout (11.8 vs. 28.2 minutes). As shown in Figure~\ref{fig:failure-counts}, this severe E2E inflation in Eager is largely driven by out-of-memory (OOM) failures and task timeouts that force costly workflow retries. By proactively managing memory boundaries, \sysname{} avoids these thrashing behaviors, directly converting reliable execution into substantial throughput gains.
\begin{figure}[t]
    \centering
    \includegraphics[width=1\linewidth]{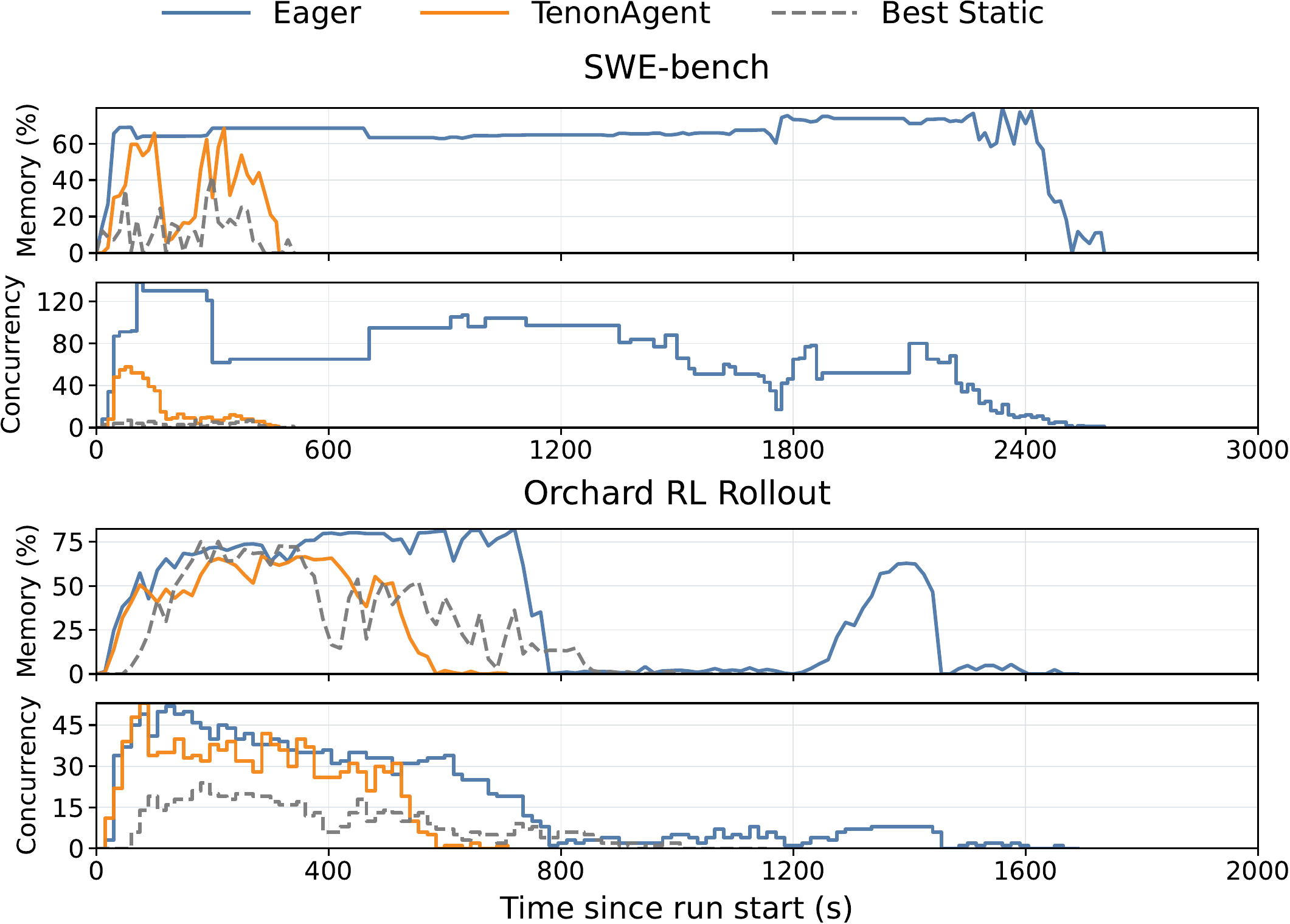}
    \caption{\sysname{} dynamically regulates memory use and execution concurrency on SW-bench and Orchard RL rollouts.}
    \label{fig:end-to-end-memory}
\end{figure}
Conversely, conservative memory policies do not guarantee efficiency. Both Agent-level and \texttt{AgentCgroup} experience extreme degradation; for instance, under 4U8G, \texttt{AgentCgroup} extends E2E makespans to 120.6 and 293.4 minutes on SWE-bench and ScienceAgentBench, respectively. In contrast, \sysname{} completes these same batches in just 21.3 and 7.9 minutes. Overall, \sysname{} yields massive absolute improvements: \textbf{up to a $37\times$ speedup over \texttt{AgentCgroup}, nearly $5\times$ over Agent-level reservations} (105.6 vs. 21.3 minutes on SWE-bench), \textbf{and roughly $2\times$ over the best-performing fixed-concurrency slots}. This confirms that rigid reservations or tool-call-aligned throttling incur excessive queuing overhead and fail to smoothly accommodate the phase transitions of long-lived agents. We provide a detailed root-cause analysis of why \texttt{AgentCgroup} encounters these extreme execution stalls in Appendix. While abundant resources naturally narrow the absolute performance gap between \sysname{} and Eager (e.g., closing to 9.3 vs. 9.6 minutes on Orchard RL at 8U16G), \sysname{} still consistently delivers superior median (P50) workflow latencies across the board. Ultimately, by effectively bounding tail failures under pressure and proactively reclaiming idle footprints, \sysname{} maximizes batch throughput while remaining highly competitive when resources are plentiful.

Figure~\ref{fig:end-to-end-memory} shows memory utilization and
concurrency over time for two workloads. Eager execution sustains high
concurrency, creating substantial memory pressure. In contrast, the
Best Static policy is conservative and often leaves memory capacity
underutilized. \sysname{} dynamically adjusts execution concurrency to
the workload's memory demand, balancing memory utilization against
concurrent growth rather than applying a fixed limit.

\subsection{Portability to Firecracker Backend}
\label{sec:eval:firecracker}

\begin{figure}
    \centering
    \includegraphics[width=1\linewidth]{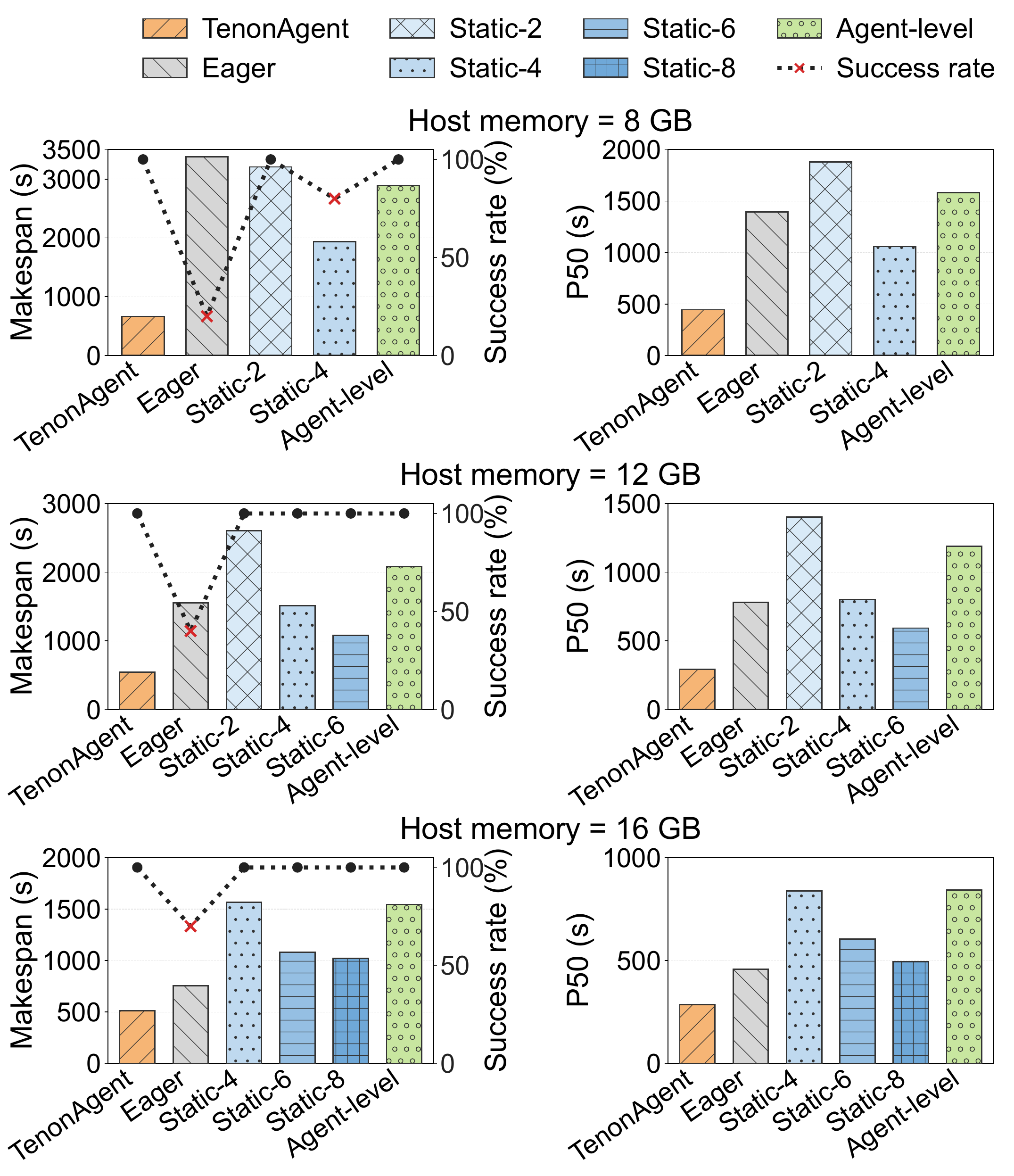}
    \caption{Performance with the Firecracker-based execution backend under different host memory budgets. The left column displays the batch makespan (bars, left axis) and workflow success rate (dotted line, right axis). The right column reports the median (P50) end-to-end workflow latency.}
    \label{fig:firecracker-metrics}
\end{figure}
To verify whether the benefits of \sysname{} generalize beyond its Docker-based implementation, we migrate the execution backend to Firecracker microVMs and evaluate it under constrained host memory budgets of 8, 12, and 16\,GB. 
As shown in Figure~\ref{fig:firecracker-metrics}, \sysname{} demonstrates consistent robustness across all configurations, achieving a 100\% attempt success rate. 
In contrast, the Eager baseline triggers severe memory contention and out-of-memory (OOM) failures, succeeding in only 20\%--70\% of attempts. 
Similarly, aggressive fixed-slot configurations trade reliability for concurrency, frequently crashing when concurrent microVMs experience overlapping memory spikes. By dynamically regulating residency and execution permissions based on global memory pressure, \sysname{} safely navigates these strict capacity limits.

Beyond simply avoiding failures, \sysname{} consistently outperforms the safe baselines in both throughput and latency. 
For a fair comparison, we isolate the best static configurations that manage to achieve 100\% success at each budget (Static-2, Static-6, and Static-8). 
Against these conservative baselines, \sysname{} achieves a $1.97\times$--$4.83\times$ speedup in batch makespan and reduces P50 end-to-end workflow latency by $1.7\times$--$4.2\times$. 
It also significantly outperforms the agent-level reservation baseline, delivering $3.02\times$--$4.36\times$ makespan reductions. 
These substantial performance gains emerge because \sysname{} actively reclaims memory from idle microVMs waiting on external LLM inference, repurposing that headroom to admit more active tool executions concurrently.

Overall, these results confirm that \sysname{}'s core design—decoupling logical agent orchestration from physical execution residency—is fundamentally agnostic to the underlying isolation mechanism.

\subsection{Offered-load Sensitivity}
\label{sec:eval:qps}

\begin{figure}[t]
    \centering
    \includegraphics[width=\linewidth]
        {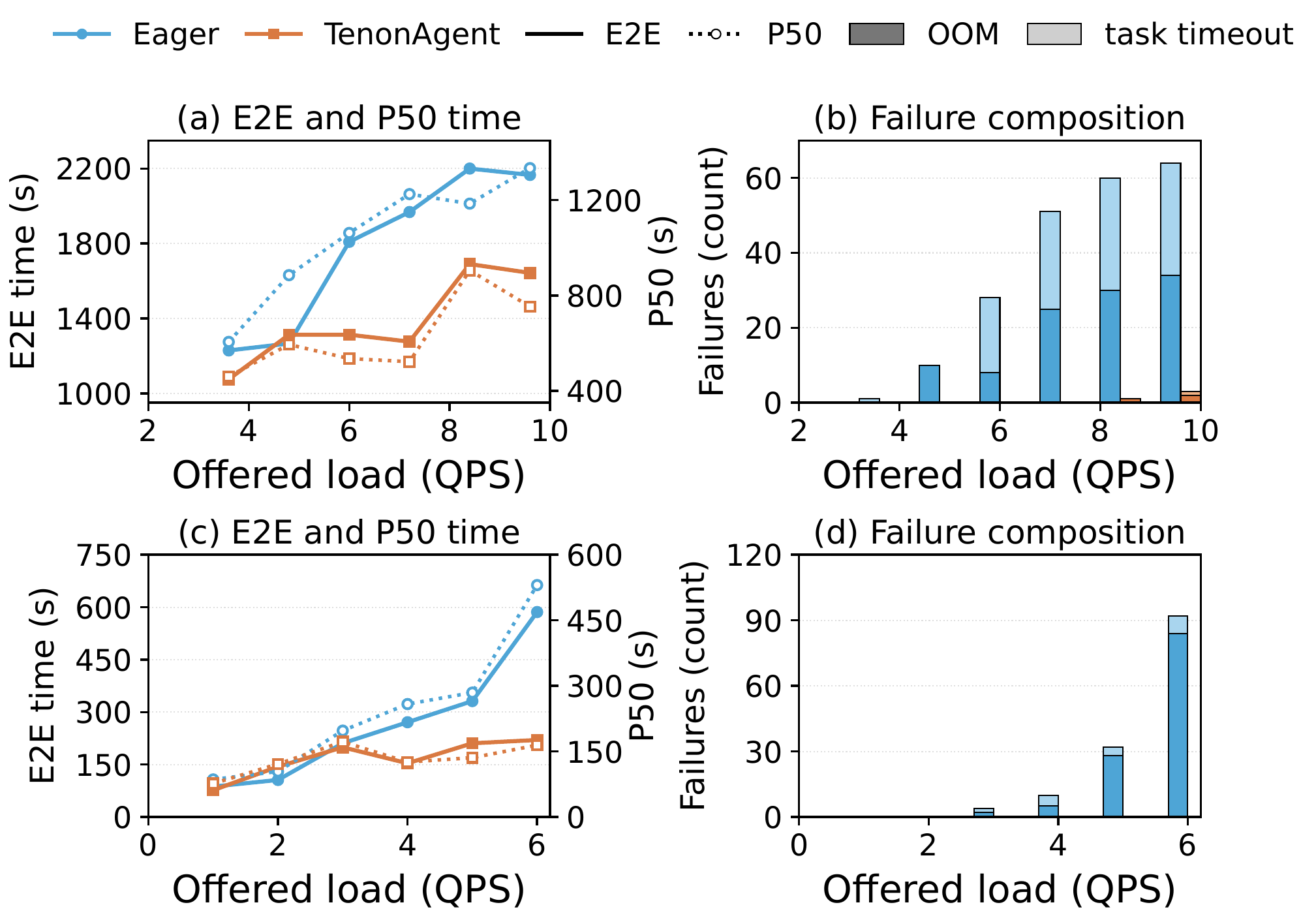}
    \caption{Sensitivity to offered-load Sensitivity. The top row evaluates bursts composed of different taks, while the bottom row evaluates an extreme duplicated-task burst. Left panels report E2E and P50 latency; right panels
    report execution failures, separated into OOM and timeout failures.}
    \label{fig:qps-sensitivity}
\end{figure}

Figure~\ref{fig:qps-sensitivity} evaluates \sysname{}'s resilience to short-duration load bursts by varying the submission rate (QPS) over a fixed 10-second window. Consequently, higher QPS increases both arrival intensity and the total workflow volume, directly driving up the \textbf{memory oversubscription ratio} (reaching up to \textbf{8.6$\times$} the physical host capacity at peak demand). This evaluates tolerance to finite-duration bursts rather than sustainable steady-state throughput. The top row of Figure~\ref{fig:qps-sensitivity} demonstrates a realistic \emph{mixed-workflow} setting, where diverse execution trajectories prevent peak memory demands from aligning. At low QPS (low oversubscription), \sysname{} and eager execution achieve comparable latencies, indicating minimal scheduling overhead. However, as burst intensity scales and the oversubscription ratio surges, eager execution degrades significantly, suffering severe latency spikes and frequent out-of-memory (OOM) and timeout failures. In contrast, \sysname{} maintains stable end-to-end and P50 latencies while minimizing failures, proving the value of resource-aware control under realistic bursts. The bottom row introduces an adversarial \emph{duplicated-workflow} stress test, concurrently submitting identical workflows to force their high-memory phases to synchronize. This alignment creates an extreme local oversubscription scenario. Without control, eager execution collapses under this pressure, exhibiting soaring tail latencies and massive OOM failures. Conversely, \sysname{} actively smooths this synchronized demand via admission and execution throttling, maintaining low latency and avoiding failures entirely across all tested loads.

Ultimately, \sysname{} delivers robust performance under both diverse and
worst-case synchronized bursts. It imposes negligible overhead during light
traffic while preventing OOM-driven degradation even when the memory
oversubscription ratio, defined as the sum of offline-profiled peak memory
footprints of all submitted workflows divided by the host memory capacity, is
pushed to an extreme \textbf{8.6$\times$}.

\begin{figure}[t]
    \centering
    \includegraphics[width=\linewidth]
        {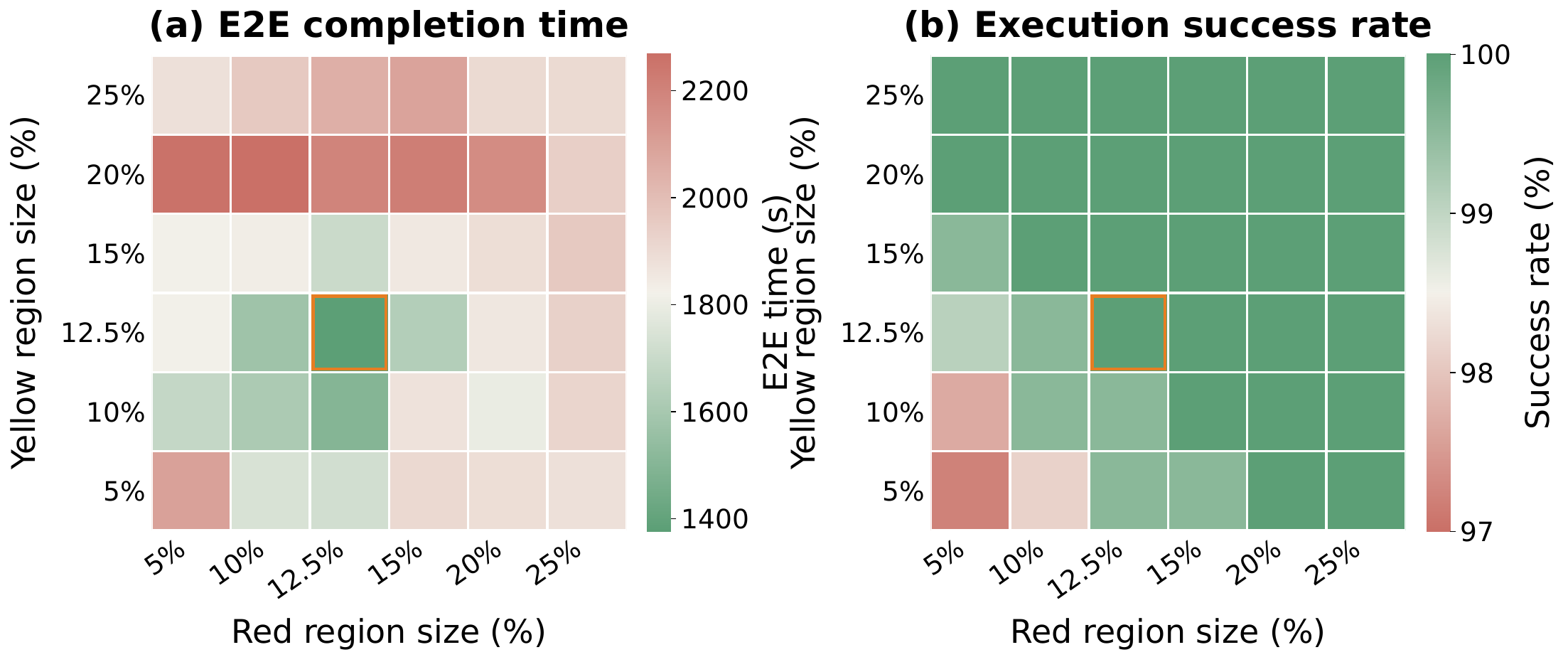}
    \caption{Sensitivity to \textsc{red}- and \textsc{yellow}-region sizes: (a) E2E completion time and (b) execution success rate. Green indicates better outcomes (lower time or 100\% success rate), while red indicates worse outcomes. The reference configuration (12.5\% for both) is outlined in orange.}
    \label{fig:memory-region-sensitivity}
\end{figure}

\subsection{Memory-Region Sizing Sensitivity}
\label{sec:eval:region-sizing}

We examine how the relative sizes of the \textsc{red} and \textsc{yellow} memory regions affect workload performance and execution reliability. Unlike ordered boundary thresholds, these parameters are defined as percentages of the total memory budget and are varied independently. Figure~\ref{fig:memory-region-sensitivity} illustrates a parameter sweep with region sizes ranging from 5\% to 25\%. The reference configuration (\textsc{red} = 12.5\%, \textsc{yellow} = 12.5\%) is outlined in orange.

\paragraph{Reliability requires sufficient emergency headroom.}
Figure~\ref{fig:memory-region-sensitivity}b demonstrates that \sysname{} maintains a 100\% execution success rate across most configurations. However, reliability drops when both the \textsc{red} and \textsc{yellow} regions are aggressively small (e.g., \textsc{red} $\le$ 5\%, \textsc{yellow} $\le$ 10\%). In the \textsc{red} state, the kernel actuator suspends active sandboxes to halt memory growth. If the \textsc{red} region is too small, the system lacks the physical headroom to absorb instantaneous allocation spikes (overshoots) that occur between the controller's observation and the physical enforcement, ultimately resulting in OOM failures.

\paragraph{Performance trade-offs of preventive and reactive control.}
Figure~\ref{fig:memory-region-sensitivity}a reveals distinct performance penalties for over-provisioning either region. When the \textsc{yellow} region is excessively large (e.g., 20\%--25\%), E2E completion time increases significantly, exceeding 2000 seconds. A large \textsc{yellow} zone makes the system overly conservative, triggering premature preventive reclamation of idle sandboxes. This destroys warm environments that could have been reused, introducing severe sandbox reconstruction overhead. Conversely, an oversized \textsc{red} region (e.g., 20\%--25\%) stretches completion time by prematurely suspending active growth contributors, unnecessarily reducing execution concurrency before true memory pressure materializes.

The reference configuration (12.5\% for both regions) hits the empirical ``sweet spot''. It provisions a \textsc{yellow} region large enough to give background sandbox teardowns sufficient time to complete, paired with a \textsc{red} region deep enough to safely buffer active allocation overshoots. This dual-region balance safely maximizes concurrency and achieves the lowest workload makespan.

\subsection{Evaluation of Memory Prediction}
\label{sec:eval:predictor}

We compare three action-level predictors: \textit{Static} (uses only action semantics), \textit{History-only} (reuses the preceding action's memory demand), and \textit{Current} (combines semantics with causal execution history). As shown in Table~\ref{tab:predictor-accuracy}, relying on isolated signals degrades prediction quality. The \textit{Static} predictor suffers from massive numerical errors (e.g., a P90 APE of 442.3\%) and low precision (52.0\%), generating many false positives by over-predicting heavy actions. Conversely, \textit{History-only} reduces overall numerical error but its heavy-action recall plummets to 48.8\%, proving that past demand alone cannot reliably anticipate sudden memory spikes.

Our \textit{Current} predictor synthesizes both complementary signals, dominating all metrics. Compared to Static, it halves the Mean Absolute Error (MAE) and slashes P90 APE by 80\%, while achieving the best balance of heavy-action precision (77.3\%) and recall (82.9\%). We further demonstrate how this precise demand estimation translates to end-to-end system reliability and latency improvements in Section~\ref{sec:eval:ablation}.

\begin{table}[t]
\centering
\caption{Direct evaluation of memory-demand predictors. Lower is better
for MAE and P90 APE; higher is better for heavy-action precision and
recall. Heavy actions have measured peak memory above 1,024\,MiB.}
\label{tab:predictor-accuracy}
\resizebox{\linewidth}{!}{%
\begin{tabular}{lrrrr}
\toprule
& \multicolumn{2}{c}{Numerical prediction}
& \multicolumn{2}{c}{Heavy classification} \\
\cmidrule(lr){2-3}
\cmidrule(lr){4-5}
Method
& MAE (MiB)
& P90 APE (\%)
& Precision (\%)
& Recall (\%) \\
\midrule
Static
& 198.6
& 442.3
& 52.0
& 80.0 \\

History-only
& 154.3
& 175.2
& 50.7
& 48.8 \\

Current
& \textbf{93.9}
& \textbf{88.8}
& \textbf{77.3}
& \textbf{82.9} \\
\bottomrule
\end{tabular}%
}
\end{table}



\subsection{Online Agent Evaluation}

Unlike replay-based experiments, this evaluation runs live,
LLM-driven agents on SWE-bench and ScienceAgentBench. All online runs use a
single workload level on a 4U8G VM. We evaluate two Qwen3.8 configurations,
Qwen3.8-Flash~\cite{qwen38flash} and Qwen3.8-Max~\cite{qwen38max0902}, served by the Alibaba Cloud Qianwen platform.
For each benchmark and model configuration, we compare eager execution
(\emph{w/o TenonAgent}) with execution under \sysname{} (\emph{w/ TenonAgent})
using the same task inputs, agent implementation, LLM configuration, resource
budget, and workflow arrival schedule. Since online LLM-driven execution is
inherently variable, each configuration is repeated five times; bars report the
mean and error bars show variation.

\begin{figure}[t]
    \centering
    \includegraphics[width=\linewidth]
        {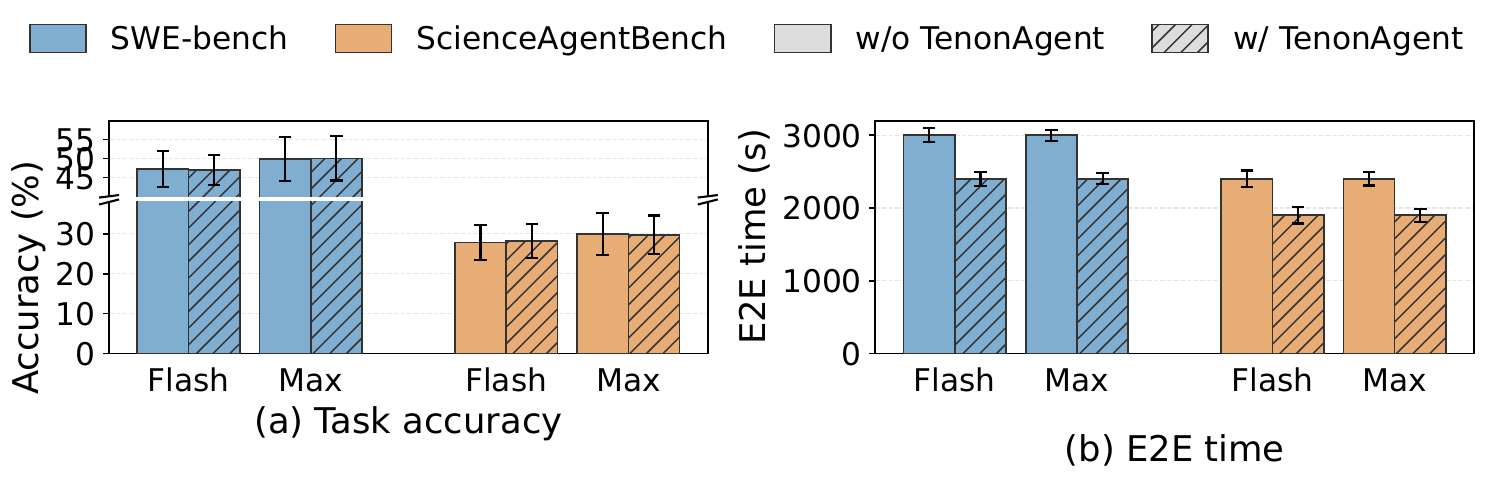}
    \caption{Online evaluation on SWE-bench and ScienceAgentBench under
    low and moderate workload levels: (a) task accuracy and (b)
    end-to-end execution time, comparing eager execution
    (\emph{w/o TenonAgent}) with \sysname{} (\emph{w/ TenonAgent}).}
    \label{fig:online_accuracy_time}
\end{figure}

Figure~\ref{fig:online_accuracy_time} reports task accuracy and end-to-end
execution time. Across both benchmarks and both Qwen3.8 configurations,
\sysname{} preserves task accuracy, with w/ and w/o TenonAgent bars closely
matched in all cases. In contrast, end-to-end time consistently decreases under
\sysname{}. These results suggest that
\sysname{} improves execution efficiency in online setting without an observable loss in task accuracy.
We further examine restart-safe coverage and lifecycle overhead. Table~\ref{tab:teardown-results} reports averages across five
runs. SWE-bench executes 1,952.8 actions on average, with 93.4\% restartable
boundaries; ScienceAgentBench executes 536.2 actions on average, with 100\%
restartable boundaries. \sysname{} performs 396.8 and 124.4 teardowns on
average, reclaiming 33.6\,GiB and 8.6\,GiB of memory, respectively, with
cumulative lifecycle overheads of 805.0\,s and 530.4\,s. These results show
broad restart-safe coverage and substantial memory reclamation, while
quantifying the teardown and reconstruction cost.

\begin{table}[t]
\centering
\small
\caption{Restart-safe coverage and teardown overhead in higher-load agent runs.}
\label{tab:teardown-results}
\begin{tabular}{@{}lcc@{}}
\toprule
Metric & SWE-bench & ScienceAgentBench \\
\midrule
Actions                    & 1,952.8 & 536.2 \\
Restartable boundaries (\%) & 93.4\% & 100\% \\
Teardowns                  & 396.8 & 124.4 \\
Memory reclaimed (GiB)     & 33.6 & 8.6 \\
Cumulative overhead (s)    & 805.0 & 530.4 \\
\bottomrule
\end{tabular}
\end{table}

\begin{table*}[t]
    \centering
    \small
    \renewcommand{\arraystretch}{1.1}
    \setlength{\tabcolsep}{5pt}
    \caption{Ablation results under an 8\,GB memory budget.
    Each variant removes one mechanism or replaces phase-level
    control with agent-level control while retaining the
    remaining mechanisms. Parentheses indicate latency
    increases relative to the full system.
    Out-of-memory and sandbox timeout columns report event counts.}
    \label{tab:ablation-8gb}

    \resizebox{\textwidth}{!}{%
    \begin{tabular}{@{}lcccccc@{}}
        \toprule
        Configuration
        & OOM events
        & Sandbox timeouts
        & End-to-end time (s)
        & P50 latency (s)
        & Peak concurrency
        & Peak memory (MB) \\
        \midrule

        \sysname{}
        & \textbf{0}
        & \textbf{0}
        & \textbf{1276}
        & \textbf{522}
        & 34
        & 7415 \\

        \textit{W/O admission control}
        & 1
        & 0
        & 1489 {\scriptsize (+16.7\%)}
        & 721 {\scriptsize (+38.1\%)}
        & 42
        & 7937 \\

        \textit{W/O memory-growth control}
        & 3
        & 0
        & 1841 {\scriptsize (+44.3\%)}
        & 842 {\scriptsize (+61.3\%)}
        & 27
        & 7978 \\

        \textit{Always resident}
        & 2
        & 1
        & 1845 {\scriptsize (+44.6\%)}
        & 1103 {\scriptsize (+111.3\%)}
        & 43
        & 8034 \\
        \hline
        \textit{W/ static predictor}
        & 0
        & 0
        & 2441 {\scriptsize (+91.3\%)}
        & 1212 {\scriptsize (+132.2\%)}
        & 20
        & 5414 \\
        \textit{W/ history-only predictor}
        & 2
        & 1
        & 1706 {\scriptsize (+33.7\%)}
        & 836 {\scriptsize (+60.2\%)}
        & 38
        & 7965 \\
        \bottomrule
    \end{tabular}%
    }
\end{table*}
\subsection{Ablation Study}
\label{sec:eval:ablation}

We evaluate the individual contributions of admission control, memory-growth control, dynamic sandbox residency, and the demand predictor under an 8\,GB memory budget. Starting from the full \sysname{} system, we construct five degraded variants. Table~\ref{tab:ablation-8gb} summarizes the performance, reliability, and concurrency metrics for each configuration.

\textbf{Core Runtime Mechanisms.}
The full system delivers the best overall performance (1276\,s E2E, 522\,s P50) with zero OOM or timeout failures. Disabling proactive admission control allows unvetted memory growth, increasing E2E time by 16.7\% and triggering an OOM event. Removing kernel-level memory-growth control is even more detrimental, increasing E2E time by 44.3\% and causing three OOMs, as unexpected allocations during execution go unthrottled. Furthermore, an \textit{always resident} policy---which disables restart-safe teardown---severely penalizes performance (+111.3\% P50 latency). This confirms that static residency inevitably strands memory, whereas dynamic lifecycle decoupling is essential for surviving strict capacity limits.

\textbf{Efficacy of the Predictor.}
To isolate the value of \sysname{}'s contextual prediction, we evaluate two degraded baselines. A \textit{static predictor} heavily overestimates memory demands, enforcing a conservative admission policy that starves the system: peak concurrency drops to 20 and memory remains underutilized (5414\,MB). While this reduces OOMs, excessive queuing inflates P50 latency. Conversely, a \textit{history-only predictor} frequently misses sudden memory spikes. It admits actions too aggressively, plunging the system into severe memory contention that causes 2 OOMs and increases E2E time by 33.7\% due to recovery overheads. Ultimately, \sysname{}'s full predictor achieves the optimal balance, precisely gauging demand to safely maximize concurrency (34) without triggering failures or inflating latency.

\section{Related Work}
\label{sec:related-work}
\myparagraph{LLM Agents and Agentic Serving}
Recent agent systems transform LLM calls into long-running, stateful workflows interacting with diverse external environments~\cite{yao2022webshop,liu2023agentbench,yang2024sweagent,xu2024crab,xie2024osworld,zhou2024webarena}, supported by frameworks providing multi-agent and tool-use abstractions~\cite{langchain,langgraph,autogen,crewai2025,llamaindex,gao2024agentscope}. Emerging research treats these executions as schedulable programs: Alto~\cite{raghavan2025altoorchestratingdistributedcompound}, Autellix~\cite{luo2025autellixefficientservingengine}, Helium~\cite{helium2026agentic}, and Parrot~\cite{lin2024parrot} orchestrate distributed compound systems, while Continuum~\cite{li2025continuum} and ThunderAgent~\cite{kang2026thunderagent} optimize GPU-side KV-cache retention for multi-turn interactions. However, agentic workloads also severely stress CPU-side memory and tool orchestration~\cite{agenticcpu2025}. Unlike systems predominantly optimizing distributed routing or accelerator inference, \sysname{} addresses the orthogonal bottleneck of host-side memory oversubscription. It explicitly governs the local CPU-side execution substrate---managing how and when sandboxes consume memory.

\myparagraph{LLM Serving and Sandbox Runtimes}
Mainstream LLM serving engines maximize accelerator throughput via iteration-level scheduling and memory management~\cite{yu2022orca,kwon2023efficient,agrawal2024sarathi,zhong2024distserve,zheng2024sglang}. While essential for inference, they omit the CPU-side environments (e.g., reconstructable workspaces and tool outputs) required by agents~\cite{cheng2026llminsandbox}. To isolate these substrates, runtimes like Firecracker and gVisor provide high-density containment~\cite{akkus2018sand,agache2020firecracker,ao2022faasnap,li2022rund,gvisor,nsjail}. For stateful workflows, tools like CRIU~\cite{andrijauskas2024criu} and DeltaBox~\cite{dong2026deltabox} employ OS-level checkpoint/restore (C/R) to freeze process states. However, these VMs and containers preserve the rigid identity of the execution instance. In contrast to traditional C/R mechanisms that rely on heavy memory-heap snapshots to preserve physical state, \sysname{} virtualizes execution residency at the semantic layer by decoupling the logical agent from its physical sandbox.


\section{Discussion and Limitations}
\label{sec:discussion}

\paragraph{Scope of host-resource management.}
\sysname{} primarily targets host memory capacity, which is non-compressible
and often determines failure under oversubscription. Lifecycle limits help
bound reconstruction overhead, but \sysname{} does not explicitly model CPU
contention, memory bandwidth, or application-level I/O. Although Appendix
provides a preliminary I/O analysis, these factors may dominate before memory
is exhausted in data-intensive workloads. Extending residency virtualization
to multi-resource co-scheduling remains future work.

\paragraph{Sensitivity to workflow dynamics.}
Residency virtualization relies on idle intervals during external LLM reasoning. If inference latency drops significantly, shrinking reclamation windows could cause sandbox reconstruction overheads to dominate. Furthermore, \sysname{} assumes observable action boundaries within dedicated sandboxes; supporting uncoordinated multi-tenant workloads would require finer-grained intra-sandbox isolation.

\section{Conclusion}
\label{sec:conclusion}

In this paper, we present \sysname{}, a runtime system that virtualizes sandbox
execution residency by decoupling a logical agent's lifetime from its underlying
physical sandbox. By coordinating boundary-scale residency decisions with
kernel-assisted execution pacing, \sysname{} reduces observed out-of-memory
failures and delivers up to a $5.5\times$ speedup over unmanaged execution
under constrained host budgets. In our evaluated workloads and parameter
settings, it achieves a 100\% task success rate under bursts reaching an
extreme \textbf{$8.6\times$ memory oversubscription ratio}, showing that
physical residency can be elasticized to improve high-density agent serving
while preserving sandbox-based isolation.
\sysname{} provides a practical foundation for
memory-oversubscribed agentic AI infrastructure.

\bibliographystyle{ACM-Reference-Format}
\bibliography{paper}
\clearpage

\appendix

\section{Parameter Sweep}
\label{sec:eval:sensitivity}

\begin{figure}
    \centering
    \includegraphics[width=\linewidth]{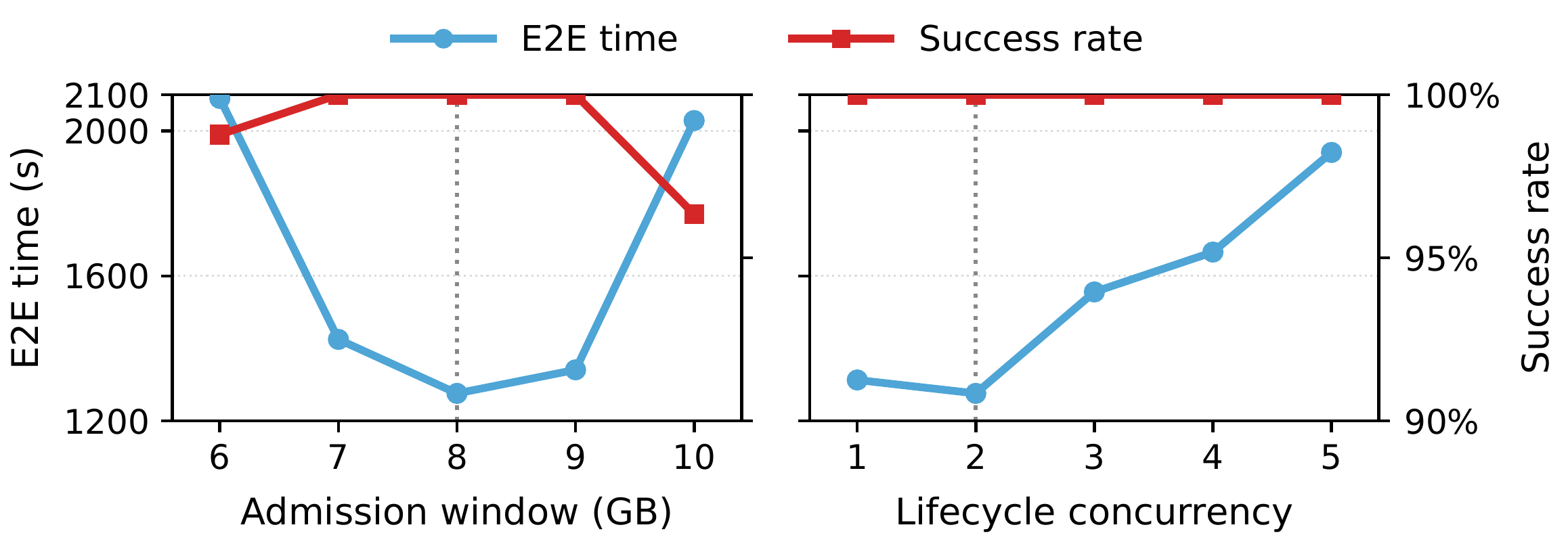}
    \caption{Sensitivity to (a) the upper bound of the computed admission
    window and (b) the lifecycle concurrency limit. Each parameter is varied
    independently while all other mechanisms remain enabled.}
    \label{fig:parameter-sensitivity}
\end{figure}

Figure~\ref{fig:parameter-sensitivity} evaluates two parameters of the
user-space boundary-scale scheduler. We vary one parameter at a time while
keeping all other mechanisms enabled. The reference configuration uses an
8\,GB upper bound for the computed admission window and a lifecycle
concurrency limit of two. Under this configuration, the workload completes
in 1276\,s without OOM or sandbox-timeout events.

\paragraph{Admission-window upper bound.}
The admission window is computed from the predicted resource demand of
upcoming actions, as described in \S\ref{sec:design:user-scheduler}. The
parameter studied here is the upper bound imposed on this computed window,
rather than a fixed reservation assigned uniformly to every action. When the
upper bound is below 8\,GB, admission becomes overly conservative and limits
useful overlap among actions, resulting in longer completion time. Increasing
the upper bound beyond 8\,GB does not improve completion time and eventually
reduces the observed success rate because more predicted memory demand is
allowed to accumulate concurrently. The 8\,GB setting therefore provides the
best balance between execution concurrency and protection against concurrent
memory pressure for this workload.

\paragraph{Lifecycle concurrency limit.}
Sandbox start, stop, and rematerialization operations involve substantial I/O
and share limited bandwidth and host-side resources. The lifecycle
concurrency limit therefore serves as a rate limiter, bounding the number of
lifecycle transitions that may be active simultaneously. A limit of two
achieves the lowest E2E completion time in the sweep. Increasing the limit
beyond two creates additional overlap but does not increase effective
lifecycle throughput; instead, contention grows and completion time
increases. Higher limits also introduce sandbox-timeout events. These results
indicate that lifecycle parallelism should be bounded by the available I/O
bandwidth and related host resources, rather than maximized independently of
the underlying service capacity.

\section{Runtime Control Parameters}
\label{app:runtime-control-parameters}

Table~\ref{tab:full_arm_control_params} gives the default
configuration for the runtime execution-control specification
in Section~\ref{sec:design:kernel-enforcement}.
The algorithm and its accompanying definitions determine the
decision order; this table supplies parameter values rather
than an alternative policy.

\paragraph{Control-iteration order.}
Each control iteration performs the following operations:
(1) remove completed sandboxes and register newly observed ones;
(2) collect footprint and CPU-time observations;
(3) update estimates and states only for valid sample intervals;
(4) compute projected demands and pressure;
(5) update the concurrency budget;
(6) select the complete permitted set; and
(7) publish permissions and update denial episodes.
All decisions within an iteration use the same observation
snapshot and control timestamp.

A completed sandbox is removed from all candidate and permitted
sets, and its estimator and denial timer are discarded.
A newly registered sandbox has no valid growth estimate,
starts in Growing, has no active denial timer, and uses
$d_{\mathrm{boot}}$ until its first valid sample.
Its execution permission is initially \textsc{deny} until the
controller publishes its first decision.
If that decision is \textsc{deny}, the denial timer starts at
that iteration.

\paragraph{Controller initialization.}
The initial permitted set is empty, the initial concurrency
budget is $b_0$, and the previous pressure is initialized to
Green for pressure-transition bookkeeping.
The Green stability timer starts when the first Green decision
is made and is cleared whenever pressure is non-Green.

While pressure remains Green, each completed
$T_{\mathrm{stable}}$ interval increments $b$ once.
If a delayed control iteration spans multiple complete
stability intervals, the controller applies one increment
per complete interval, capped at $b_{\max}$, and retains the
remaining fractional interval.
A pressure increase applies multiplicative decrease once
per observed transition, not once per control iteration.
A decrease in pressure does not directly change $b$.

\paragraph{Measurement validity.}
The footprint baseline is refreshed at every sample, including
samples collected while execution is denied.
Consequently, the first subsequent valid growth sample uses
the immediately preceding footprint observation rather than
accumulating changes over the entire denial episode.
An interval containing a policy-permission change, or having
no increase in accounted CPU time, is invalid for updating
the growth estimate.
Such an interval does not supply evidence of reduced growth.

There is no separate inactive-footprint cutoff and no
additional re-trust timer.
The near-zero band affects the growth-state label, not
headroom accounting: any strictly positive predicted demand
consumes a contributor slot and is included in the aggregate
projection.

\begin{table}[t]
\centering
\caption{Default parameters for the runtime execution-control
specification. Runtime growth accounting is separate from
action-level admission accounting.}
\label{tab:full_arm_control_params}
\footnotesize
\setlength{\tabcolsep}{3pt}
\renewcommand{\arraystretch}{1.08}
\begin{tabularx}{\columnwidth}{@{}p{0.34\columnwidth}X@{}}
\toprule
\textbf{Parameter} & \textbf{Default value / definition} \\
\midrule

Footprint &
$m_c=\texttt{memory.current}_c+
\texttt{memory.swap.current}_c$. \\

Sampling interval $\Delta$ &
$100\,\mathrm{ms}$. \\

Controller period &
$100\,\mathrm{ms}$. \\

Scheduler update polling &
$100\,\mathrm{ms}$; not a hard bound on actuation latency. \\

Projection horizon $T$ &
$1000\,\mathrm{ms}$. \\

EWMA factor $\alpha$ &
$0.35$; the first valid raw sample initializes the estimate. \\

Resident band $\epsilon$ &
$4\,\mathrm{MiB/s}$;
Resident iff $|\hat g_c|\le\epsilon$. \\

Bootstrap demand $d_{\mathrm{boot}}$ &
$16\,\mathrm{MiB}$ over the projection horizon, used only
until a valid growth estimate exists. \\

Safety margin $S$ &
$256\,\mathrm{MiB}$. \\

Usable headroom $H$ &
$\max(\texttt{MemAvailable}-S,0)$. \\

Green rule &
Allow every live sandbox. Growth-window, contributor-count,
and denial-waiting restrictions are bypassed. \\

Yellow budget &
$b_0=1$, $b_{\max}=4$.
At most $b$ positive-demand sandboxes may be permitted,
with aggregate demand at most $H$. \\

Multiplicative decrease &
$b\leftarrow\max(1,\lfloor b/2\rfloor)$ once on each
observed increase in pressure severity. \\

Additive increase &
Increment $b$ by one per
$T_{\mathrm{stable}}=5000\,\mathrm{ms}$
of continuous Green operation, capped at $b_{\max}$. \\

Red-drain budget &
At most one positive-demand sandbox in the complete
permitted set; aggregate demand must remain at most $H$. \\

Minimum denial wait $D_{\min}$ &
$1000\,\mathrm{ms}$ before a denied sandbox is eligible
for admission in Yellow or Red-drain. Green overrides it. \\

Overdue threshold $D_{\max}$ &
$3000\,\mathrm{ms}$. Eligible sandboxes reaching this
duration are checked before non-overdue candidates;
all admission constraints still apply. \\

State preference &
For non-overdue candidates:
Resident, then Cooling, then Growing/PausedGrowing;
smaller projected demand and stable identifier break ties. \\

Pressure hysteresis &
No separate threshold hysteresis.
Denial waiting and the Green stability timer have the
distinct roles defined in the text. \\

\bottomrule
\end{tabularx}
\end{table}

\section{Root-Cause Analysis of \texttt{AgentCgroup} Degradation}
\label{sec:appendix-agentcgroup-analysis}

As observed in Section~\ref{sec:eval:e2e}, the \texttt{AgentCgroup} baseline prevents catastrophic out-of-memory (OOM) crashes but suffers from extreme end-to-end time inflation (e.g., extending ScienceAgentBench makespan to 293.4 minutes under 4U8G). This section details the system-level mechanisms driving this severe degradation.

The core design of \texttt{AgentCgroup} relies on pre-execution profiling to statically classify tasks into ``High'' and ``Low'' memory demand categories based on a configurable ratio. It then enforces these classifications using Linux cgroup v2 boundaries. For tasks categorized as ``High,'' the \texttt{memory.high} limit is left unbounded, allowing them to freely consume available memory. Conversely, tasks classified as ``Low'' are subjected to a rigid, hardcoded \texttt{memory.high} threshold (e.g., set to 400\,MB in our evaluation).

Through extensive parameter tuning, we found that this static categorization fundamentally fails to generalize across the dynamic, multi-turn nature of agentic workflows. The severe latency inflation stems from the OS-level penalty of misconfiguration, which forces a binary dilemma:
\begin{itemize}
    \item \textbf{Threshold Too High (Under-throttling):} If the ``Low'' threshold is relaxed to accommodate diverse tasks, it effectively neutralizes the flow control. The system acts similarly to the Eager baseline, failing to prevent overlapping memory spikes and ultimately crashing under tight global memory budgets.
    \item \textbf{Threshold Too Low (Severe Thrashing):} When a restrictive threshold (e.g., 400\,MB) is strictly enforced, any agent operation (such as loading a dataset or compiling code) that legitimately requires more memory will repeatedly hit the \texttt{memory.high} limit. Unlike \texttt{memory.max} (which triggers the OOM killer), hitting \texttt{memory.high} forces the kernel to synchronously throttle the offending processes. The kernel aggressively forces these tasks into \emph{direct memory reclaim}, spending CPU cycles on continuously swapping pages in and out.
\end{itemize}

Consequently, the agent's tool execution essentially stalls. During our evaluation, we empirically observed severe memory I/O saturation and sustained page-fault storms (thrashing) coinciding with these stalls, as tasks advanced at a glacial pace due to constant block I/O wait and kernel reclaim overhead. This confirms that static, configuration-heavy \texttt{memory.high} thresholds are fundamentally ill-suited for the unpredictable phase transitions of LLM agents.

\textbf{This fundamental limitation of reactive cgroup mechanisms directly motivated \sysname{}'s architectural shift to \texttt{sched\_ext}.} While cgroups manage memory capacity reactively---punishing workloads with severe thrashing or OOM kills upon hitting boundaries---\sysname{} leverages \texttt{sched\_ext} to proactively govern the \emph{rate} of memory growth. By intercepting CPU scheduling, \sysname{} smoothly paces the execution of aggressively allocating sandboxes \emph{before} they trigger kernel-level memory reclaim. This approach completely bypasses the binary dilemma of cgroups, effectively preventing page-fault storms while maintaining safe, high-density concurrent execution.

\section{Restart-Safe Recovery under a Restricted Execution Model}
\label{app:state-contract}

This appendix specifies the contract for safely releasing and reconstructing sandboxes at action boundaries. Unlike OS-level checkpointing (e.g., CRIU), this mechanism operates semantically: it captures explicit persistent state between actions rather than freezing arbitrary running processes.

\subsection{Supported Execution Model \& Eligibility}

\sysname{} evaluates recoverability strictly \emph{between} actions. To qualify for restart-safe teardown, a sandbox must satisfy the following execution constraints:

\begin{itemize}
    \item \textbf{Decoupled Logical State:} The controller manages all workflow progress, LLM context, and conversation history. This state must not depend on sandbox memory.
    \item \textbf{Stateless Action Processes:} Each action runs as a fresh process. Subsequent actions cannot rely on active background daemons, open sockets, interactive shell sessions, or live interpreter heaps created by previous actions.
    \item \textbf{Explicit Filesystem Persistence:} All cross-action mutable state must be confined to designated persistent directories. The rest of the environment is considered disposable and reconstructible from a base image.
    \item \textbf{Strict Termination:} The current action and all its descendant processes must have completely terminated, with their outcomes fully recorded.
\end{itemize}

If a workflow violates these constraints (e.g., spawns a required background server), the controller conservatively aborts recovery and keeps the sandbox resident.

\subsection{Persistence and Teardown Protocol}

When a sandbox is marked for dematerialization, the controller executes a strictly ordered teardown protocol:

\begin{enumerate}
    \item \textbf{Block Dispatch \& Validate:} Suspend new action dispatch and verify that the sandbox meets all eligibility constraints mentioned above.
    \item \textbf{Capture Artifacts:} Export all files within the designated persistent directories to durable storage.
    \item \textbf{Commit Recovery Record:} Generate and commit a recovery record to controller-managed storage. This record includes workflow IDs, execution generation, references to the persisted artifacts, and the base environment configuration.
    \item \textbf{Authorize Teardown:} The sandbox is destroyed \emph{only after} the recovery record is successfully committed. 
\end{enumerate}

\subsection{Rematerialization and Failure Handling}

To rematerialize a sandbox, the controller provisions a fresh instance from the recorded base environment, injects the committed filesystem artifacts, and verifies state integrity before dispatching the next action. 

\textbf{Failure Semantics:} 
\begin{itemize}
    \item \textbf{Before Teardown:} If any step fails prior to committing the recovery record (e.g., artifact upload fails or eligibility checks fail), the teardown is aborted and the sandbox safely remains \texttt{resident\_only}.
    \item \textbf{After Teardown:} Once the sandbox is destroyed, fallback to the original instance is impossible. If rematerialization fails, the controller must block execution and retry reconstruction from the committed durable record, rather than silently continuing with an incomplete state.
\end{itemize}

\end{document}